\documentclass[useAMS,usenatbib,usegraphicx,psfig]{mn2e}

\title[]{Photometric evolution of seven recent novae and the double
component characterizing the lightcurve of those emitting in $\gamma$-rays}
\author[]{U. Munari$^{1}$, F.-J. Hambsch$^{2}$, A. Frigo$^{2}$\\
$^{1}$INAF Astronomical Observatory of Padova, 36012 Asiago (VI), Italy\\
$^{2}$ANS Collaboration, c/o Osservatorio Astronomico, via 
dell'Osservatorio 8, 36012 Asiago (VI), Italy}

\begin{document}

\maketitle

\label{firstpage}

\begin{abstract}
The $B\,VI$ lightcurves of seven recent novae have been extensively mapped
with daily robotic observations from Atacama (Chile).  They are V1534 Sco,
V1535 Sco, V2949 Oph, V3661 Oph, MASTER OT J010603.18-744715.8, TCP
J1734475-240942 and ASASSN-16ma.  Five belong to the Bulge, one to SMC and
another is a Galactic disk object.  The two program novae detected in
$\gamma$-rays by Fermi-LAT (TCP J1734475-240942 and ASASSN-16ma) are Bulge
objects with unevolved companions.  They distinguish themselves in showing a
double-component optical lightcurve.  The first component to develop is the
{\em fireball} from freely-expanding, ballistic-launched ejecta with the
time of passage through maximum which is strongly dependent on wavelength
($\sim$1 day delay between $B$ and $I$ bands).  The second component,
emerging simultaneously with the nova detection in $\gamma$-rays and for
this reason termed {\em gamma}, evolves at a slower pace, its optical
brightness being proportional to the $\gamma$-ray flux, and its passage through
maximum {\em not} dependent on wavelength.  The fact that $\gamma$-rays
are detected from novae at the distance of the Bulge and at peak flux levels
differing by 4$\times$ seems to contradict some common belief like: only
normal novae close to the Sun are detected by Fermi-LAT, most normal novae
emit $\gamma$-rays, and they emit $\gamma$-rays in similar amounts.  The
advantages offered by high-quality photometric observations collected with
only one telescope (as opposed to data provided by a number of different
instruments) are discussed in connection to the actual local realization of
the standard filter bandpasses and the spectrum of novae dominated by
emission lines.  It is shown how, for the program novae, such high-quality
and {\em single-telescope} optical photometry is able to disentangle effects
like: the wavelength dependence of a fireball expansion, the recombination
in the flashed wind of a giant companion, the subtle presence of hiccups and
plateaus, tracing the super-soft X-ray phase, and determining the time of
its switch-off.  The non-detection by 2MASS of the progenitor excludes a giant
or a sub-giant being present in four of the program novae (V2949 Oph, V3661
Oph, TCP J18102829-2729590, and ASASSN-16ma).  For the remaining three
objects, by modelling the optical-IR spectral energy in quiescence it is
shown that V1534 Sco contains an M3III giant, V1535 Sco a K-type giant, and
MASTER OT J010603.18-744715.8 a sub-giant.
\end{abstract}


\section{Introduction}

The outburst of a nova originates from thermonuclear (TNR) runaway on the
surface of a white dwarf when material accreted from a companion reaches
critical conditions for ignition.  The accreted envelope is electron
degenerate, a fact that leads to violent mass ejection into surrounding
emptiness (if the donor is a dwarf) or into thick circumstellar material (if
the WD orbits within the wind of a cool giant companion).  The variety of
nova phenomena is further enriched by the dependence on WD mass (as the
velocity and amount of ejected material), the diffusion and mixing of
underlying WD material into the accreted envelope (outburst strength and
chemistry), the common-envelope interaction with the companion star (such as
3D morphology of the ejecta and duration of the post-TNR stable nuclear
burning), and the viewing angle (especially for highly structured ejecta
composed of bipolar flows, equatorial tori, diffuse prolate components and
winds or bow-shocks).  Bode \& Evans (2012) and Woudt \& Ribeiro (2014)
has provided extensive, recent reviews about classical novae.

Given the range of observable phenomena, a comprehensive description of a
nova would obviously benefits from the widest wavelength and epoch coverage,
including the pre-outburst properties of the progenitor.  While rampant
fields like X-ray imaging/spectra (eg.  Ness 2012), GeV $\gamma$-rays
detection (Ackermann et al.  2014) and radio high angular-resolution maps
(Chomiuk et al.  2014) are changing our understanding of novae, good
multi-band lightcurves are still an essential contribution.  Especially
during the initial optically thick phase, but also during the following
optically thin advanced decline, the ejecta and pre-existing circumstellar
matter reprocess at longer wavelengths (optical/IR) the energetic
input of phenomena developing primarily at much higher energies.  An accurate
and multi-band lightcurve of a nova can thus track and reveal a lot about
the powering engine and the physical conditions in the intervening and
reprocessing medium.  To be of the highest diagnostic value, a multi-band
lightcurve should be ($i$) densely mapped (daily), ($ii$) start immediately
after nova discovery and extend well into the advanced decline, stopped
only by Solar conjunction or limited by telescope diameter, ($iii$) pursue
the highest {\em external} photometric accuracy, i.e.  the combination of
the highest recorded flux with the most accurate transformation from the
{\em instantaneous} local photometric system to the standard one, and ($iv$)
the entire (or the bulk of the) lightcurve being obtained with a {\em single
instrument} and not be the result of the combination of sparse data from a
variety of different telescopes.  This last point is discussed in more
detail in sect.3 below.

The aim of the present paper is to present extensive $B\,VI$ lightcurves
of seven recent novae, all appearing at deep southern
declinations, well below those accessible with the Asiago telescopes that we
regularly use to follow spectroscopically novae appearing north of
$-$25$^\circ$.  The photometric observations presented in this paper have
been obtained with a robotic telescope we operate in Atacama (Chile).  The
program novae are listed in Table~1, together with their equatorial and
Galactic coordinates, spectral class (FeII or He/N) and date when their
discovery was announced.  All of them, except MASTER OT J010603.18-744715.8
appearing toward SMC, erupted within a few degrees of the Galactic center.

   \begin{table*}[!Ht]
   \centering
   \caption{The program novae.}
   \includegraphics[angle=90,width=17cm]{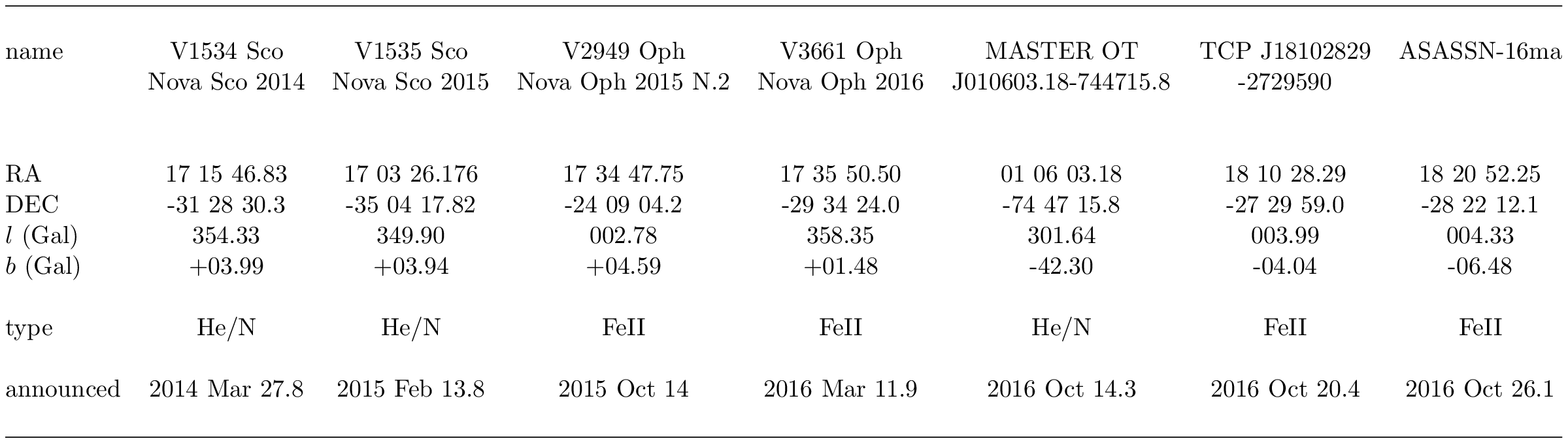}
   \end{table*}

   \begin{table*}
   \centering
   \caption{Our $B$$V$$I_{\rm C}$ photometry of the program novae. The table
   is published in its entirety electronic only.  A portion is shown here
   for guidance regarding its form and content.  Colors are given
   explicitly because, during data reduction, they are obtained independent
   from magnitudes, and are not computed one from the other.  The given
   uncertainties are the total error budgets, adding quadratically the
   Poissonian contribution on the nova to the uncertainty in the
   transformation from the instantaneous local photometric system to the
   standard one.}
   \includegraphics[width=14.5cm]{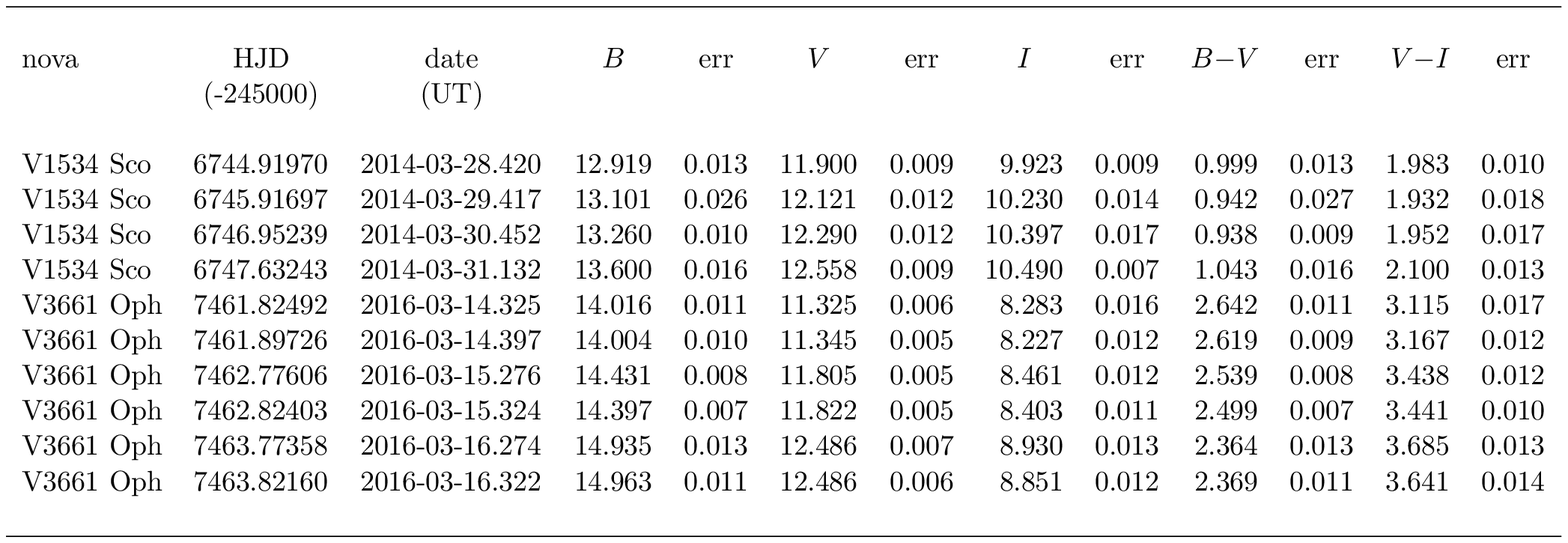}
   \end{table*}

Most of the program novae have been targeted and detected in the radio,
X-rays and/or $\gamma$-rays.  Nothing comprehensive has been yet published
on these exciting observations other than preliminary announcements on
ATels, and for only the two oldest program novae (V1534 Sco and V1535 Sco),
a comprehensive study of their near-IR spectra has been accomplished (Joshi
et al.  2015, Srivastava et al.  2015).  For none of the program novae a
detailed report on their optical properties and multi-band lightcurve has
been published so far, and thus the aim of this paper is to fill-in this gap
by providing and discussing high accuracy, daily-mapped $B\,VI$ lightcurves
for all of them.  In addition to allow by themselves some physical
discussion on the properties of the program novae, our lightcurves are meant
to provide useful support information to future studies based on other
wavelength domains.  Our photometric mapping started within one day of nova
announcement and extended until Solar conjunction set in or the nova
completed its evolution.

All times given in the paper are UT unless otherwise noted. Our photometry
is strictly tied to the Landolt (2009) system of equatorial standards, thus the
$I$ band should be properly written as $I_{\rm C}$ (Cousins' system). For
simplicity we will drop the "C" suffix in the rest of the paper, and write
the adopted photometric bands as $B\,VI$.

\section{Observations}

$B\,VI$ optical photometry of the program novae was obtained with ANS
Collaboration robotic telescope 210, located in San Pedro de Atacama, Chile. 
All novae were observed $\sim$daily for as long as Solar conjunction allowed
after their discovery.  Telescope 210 is a 40cm f/6.8 Optimized Dall-Kirkham
(ODK).  It mounts a FLI cooled CCD camera equipped with a 4k$\times$4k Kodak
16803 sensor of 9~$\mu$m pixel size.  The photometric $B\,VI$ filters are of
the multi-layer dielectric type and are manufactured by Astrodon.

Technical details and operational procedures of the ANS Collaboration
network of telescopes are presented by Munari et al.  (2012).  Detailed
analysis of the photometric performances and multi-epoch measurements of the
actual transmission profiles for all the photometric filter sets in use at
all ANS telescopes is presented by Munari \& Moretti (2012).  Data collected
on the program novae with ANS telescope  210 were ftp-transferred daily to the
central ANS server were data reduction was carried out in real time to check
on nova progress and instrument performance.  Data reduction involved all
usual corrections for bias/dark/flat/pixel map, with fresh new calibration
frames obtained regularly in spite of the highly stable conditions of the
instrumentation at the remote desert site.  Transformation from the
instantaneous local photometric system to the standard one is carried out on
all individual observations by color equations which coefficients are
$\chi^2$ calibrated against a local photometric sequence imaged together
with the nova.

The local photometric sequence is extracted from the APASS survey (Henden et
al.  2012, Henden \& Munari 2014) using the transformation equation
calibrated in Munari et al.  (2014a,b).  The APASS survey is strictly linked
to the Landolt (2009) and Smith et al.  (2002) systems of equatorial
standards.  The local photometric sequences around the program novae are
selected to fully cover the whole range of colors spanned by each individual
nova and are kept fixed during the whole observing campaign, so to ensure
the highest (internal) consistency.  Re-observing the local photometric
sequences along with the respecting novae has allowed us to refine their
magnitudes to extreme precision (well beyond the original APASS), including
pruning from the hidden presence of subtle variable stars.  These pruned and
refined local photometric sequences are freely available (e-mail the first
author) to anyone interested in using them to calibrate further optical
photometry of the novae considered in this paper.

All measurements were carried out with aperture photometry, with the
aperture radius and inner/outer radii for the sky annulus $\chi^2$-optimized
on each image so to minimize dispersion of the stars of the local
photometric sequences around the transformation equations from the local
instantaneous to the standard system.  On average, the aperture radius was
$\sim$1.0$\times$FWHM of the seeing profile, and the inner and outer radii for
the sky annulus were $\sim$3$\times$ and $\sim$4$\times$FWHM, respectively. 
Finally, colors and magnitudes are obtained separately during the reduction
process, and are not derived one from the other.  Our measurement for the
program novae are listed in Table~2, available in full only electronically. 
The quoted uncertainties are total error budgets, adding quadratically the
Poissonian contribution on the nova to the uncertainty (measured on the
stars of the local photometric sequence) in the transformation
from the instantaneous local photometric system to the standard one.

    \begin{figure}
    \includegraphics[width=82mm]{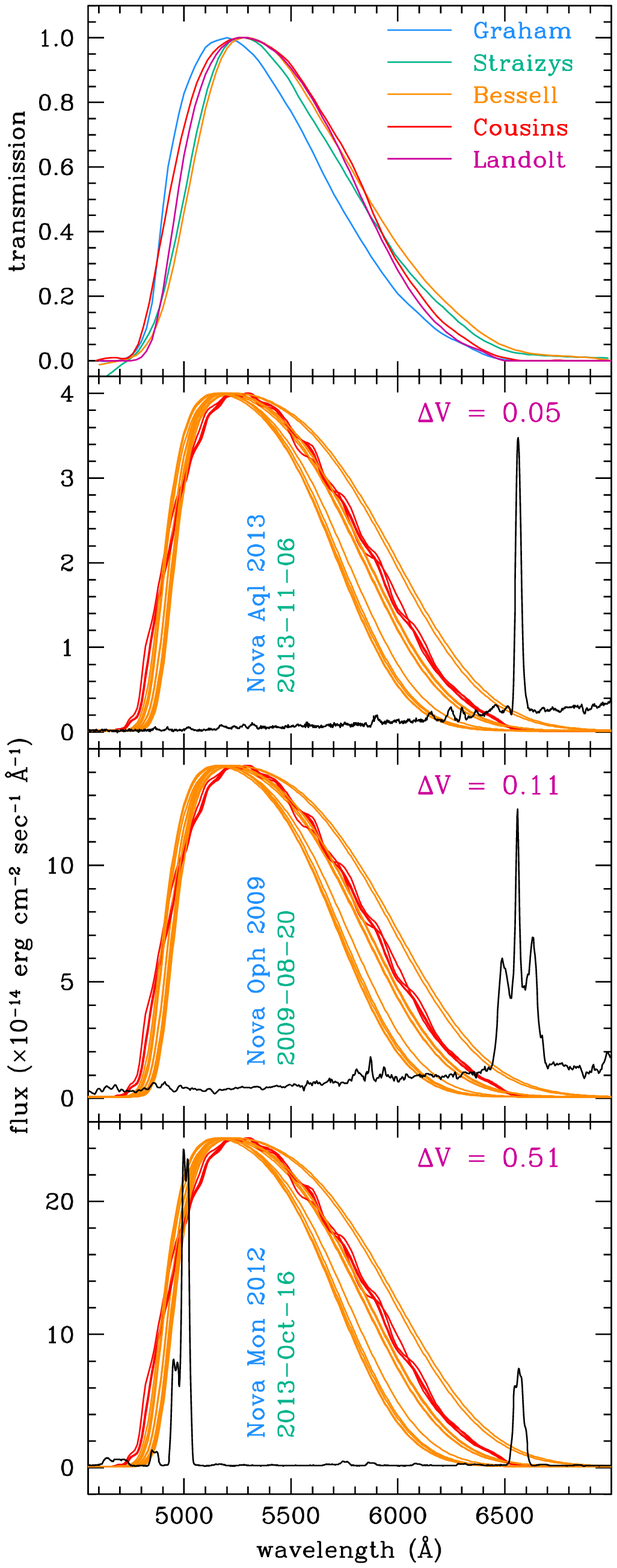}
    \caption{{\em Top panel:} natural $V$ passbands for different
    realizations of the original Johnson's $V$ band (see sect.  3 for
    details).  {\em Lower three panels:} examples of the $\Delta V$ shift
    that cannot be corrected by proper transformation to the standard
    photometric system for three types of nova spectra.  The curves are
    actual transmission profiles for $V$ band filters from different
    manufacturers, as measured in the lab with a spectrometer by Munari \&
    Moretti (2012).  See sect.3 for details.}
    \end{figure}

\section{Single-telescope vs multi-telescope lightcurves}

The advantages offered by building the lightcurve of a nova from data
provided by a single telescope (and thus not the result of the combination
of sparse data from a variety of different telescopes) are relevant, even if
frequently overlooked or not fully appreciated.  We do not refer - although
very relevant themselves - to differences in the quality of data
acquisition/reduction/calibration carried out at each telescope, nor to the
effect of difference in focal length and PSF purity in the crowded fields
where novae usually appear.  We restrict to consider here the ideal case in
which all aspects of data acquisition and reduction have been carried out
state-of-the-art and crowding is not an issue.

The spectral energy distribution of a nova is dominated by strong emission
lines, the more so as the nova declines.  While standard color-equations
(either for the all-sky and the local photometric sequence approaches) can
essentially null the differences (for normal stars and standard filter sets)
between the standard photometric system and its local instantaneous
realization (with the time- and $\lambda$-dependent atmospheric transmission
as a key component of the optical train), this is hardly so for objects
whose spectra are dominated by emission lines.

Let's consider the Landolt $V$ band for example, with similar reasoning
applicable to other bands or other photometric systems.  As discussed in
detail in Munari et al.  (2013a), much of the flux through the $V$ band
during the optically thin phase of FeII-novae comes from the [OIII] nebular
doublet.  The doublet is located on the steeply rising long-pass edge of the
$V$ band profile, where small differences in the transmission of the
photometric filters cause large deviations in the flux collected from the
nova.  Similarly, during the optically thick phase of heavily reddened novae
of both the FeII and He/N types, a non-negligible fraction of the flux
through the $V$ band comes from the H$\alpha$ emission line.  H$\alpha$ is
located at the red wing of the $V$ band, were the transmission of an actual
filter can go from null up to several \% of the peak value.

In the top panel of Figure~1 we have plot the transmission profile of the
$V$ band as locally realized by some of the most popular attempts to match
and standardize the original Johnson \& Morgan (1953) $UBV$ photometric system 
(Cousins 1980, Graham 1982, Bessell 1990, Straizys 1992, Landolt 1992). 
The differences along the whole band profile are quite obvious. Yet, proper
handling of the color-equations can essentially null such differences when
dealing with the smooth, continuum-dominated and black-body like spectral
energy distribution of normal stars.

The lower three panels of Figure~1 overplot to the spectra of three novae
the transmission profiles of the set of $V$ filters measured in the laboratory
by Munari \& Moretti (2012) with a spectrometer over the 2000 \AA\ to 1.1
$\mu$m range (so to check for either blue or red leaks).  These filters come
from the main manufacturers in the field and, prior to measurement, have
been subject to at least one year of continuous operation at the telescope
(thus exposed to large and continuous changes in barometric pressure,
temperature and humidity).  The filters are of two types.  Those following
the Bessell (1990) recipe for sandwich of Schott colored glasses (2mm of
GG495 + 3mm of BG39) are plotted in orange, the others are of the
multi-layer dielectric type and are plotted in red.  The nova spectra are
examples taken from our long term monitoring of all novae accessible with
the Asiago telescopes.  Nova Aql 2013 and Nova Oph 2009 are two heavily
reddened novae, of respectively the FeII and He/N types, as observed during
the early decline from maximum.  Nova Mon 2012 is a low reddening FeII nova
as observed during the optically thin phase, at a time when the super-soft
phase was over.

To accurately simulate actual observations of these novae with the different
$V$ filters plotted in the lower panels of Figure~1 we proceeded the following
way.  The nights when these spectra of the three novae were observed were
clear and all-sky photometric.  Several blue and red spectrophotometric
standard stars were observed at different airmass during each night.  On
the fully-extracted, but not flux-calibrated spectra of the standards, we
computed the instrumental magnitude in the $UBV$ filters which transmission
profile is fully covered by our 3200-7700 \AA\ spectra.  These instrumental
magnitudes (plus reference tabular values) were used with normal photometric
data-reduction techniques to solve the color-equations to transform from the
local to the standard photometric system.  While the $U$ and $B$ band
profiles were kept fixed to those tabulated by Landolt (1992), for the $V$
band profile we in turn adopted each one of those plotted in bottom panels
of Figure~1.

The all-sky inter-calibration of the standard stars provided stable results
at 0.01 mag level in all bands whatever the choice for the $V$ profile was. 
The $V$ magnitude derived in the same way for the novae changed instead from
one $V$ filter to another.  The range of the computed $V$ magnitudes is (cf
Figure~1) 0.05 mag for Nova Aql 2013, 0.11 mag for Nova Oph 2009, and 0.51
mag for Nova Mon 2012.  For the heavily reddened Nova Aql 2013 and Nova Oph
2009, the bluer wavelengths going through the $V$ passband contribute
essentially nothing to the recorded flux.  In such conditions, the fact that
the transmission profile of a given filter is null or is it still
transmitting something at H$\alpha$ wavelength is the reason for the
different magnitude derived for the nova.  Obviously, the wider the
equivalent width of H$\alpha$, the larger is $\Delta V$, as the comparison
of Nova Aql 2013 (e.w.(H$\alpha$)=330 \AA) and Nova Oph 2009
(e.w.(H$\alpha$)=770 \AA) clearly illustrates.  For the nebular spectrum of
Nova Mon 2012 (lowest panel of Figure~1), the line responsible for much of
the filter-to-filter differences is [OIII], which dominates (with
its e.w.(H$\alpha$)=8800 \AA) the flux going through the $V$ band. The
[OIII] doublet is located right on the steeply ascending branch of the $V$
band transmission profile, where filter-to-filter differences are the
largest.

The conclusion seems straightforward. If state-of-the-art photometry is
collected with only one telescope (always the same filters, detector,
comparison sequence, data reduction procedures), any glitch present in a
densely mapped lightcurve will probably be true, something connected to a
real change in the physical conditions experienced by the nova.  On the
contrary, if the lightcurve of a nova is built from data obtained
independently at different telescopes observing at different epochs, there
is a serious risk that any feature is an artefact caused by the mixed data
sources and it is not intrinsic to the nova.  In addition to those recorded
by ANS Collaboration, an excellent example of single-telescope lightcurves
of novae are those obtained by the SMARTS project (Walter et al.  2012).

Application of medium- and narrow-band filters to the photometry of novae
would overcome the problems caused by the mixed presence of both continuum
and emission lines within the transmission profile of broad photometric
bands, segregating the contribution of pure continuum from that of emission
lines.  The evolution of Nova Del 2013 during the first 500 days of its
eruption has been monitored by Munari et al.  (2015) simultaneously in
Landolt broad-band $B$ and $V$, Stromgren medium-band $b$ and $y$, and line
narrow-band H$\alpha$ and $[$OIII$]$ filters.  This study highlights the
great diagnostic potential of such a combined approach in carrying out the
photometry of nova outbursts.
 
\section{The program novae}

   \begin{table*}
   \centering
   \caption{Summary of some of the basic parameters for the program novae as
   derived from our lightcurves.  For TCP J18102829-2729590 and ASASSN-16ma
   the values in square brackets are computed relative to the second
   maximum, the others to the first one.  See sections on individual novae
   for further data and details.}
   \includegraphics[width=17cm]{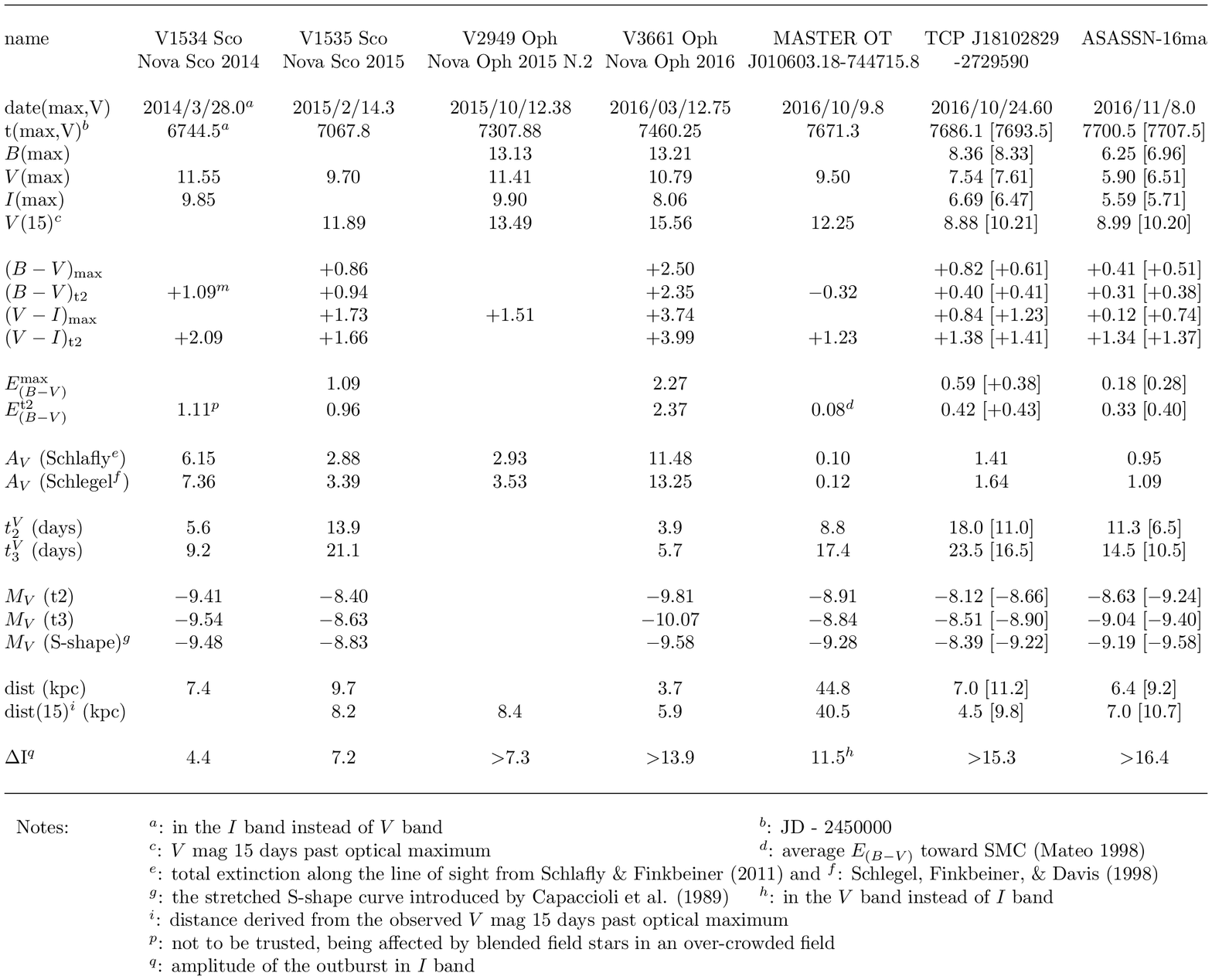}
   \end{table*}

\subsection{Decline rates, reddening and distances}

For all the program novae, the collected $B\,VI$ photometric data allows to
derive decline rates, reddening and distances with the popular methods
summarized in this section.  The results are listed in Table~3 and
in the sections below aiming to individual objects..

The characteristic rates $t_{2}^{V}$ and $t_{3}^{V}$ are the time (in days)
that a nova takes to decline in $V$ band by, respectively, 2 and 3
magnitudes below maximum brightness.  This quantity is obviously
wavelength-dependent, considering the significant color evolution presented
by a nova around maximum.  Duerbeck (2008) proposed a mean relation
$t_{3}^{V}$=1.75$\times$$t_{2}^{V}$.  For our program novae we obtain
$t_{3}^{V}$=1.54$\times$$t_{2}^{V}$ ($\sigma$=0.22).

Photometric reddening is computed by comparison with the intrinsic colors
given by van den Bergh and Younger (1987).  From a sample of well studied
novae, they derived as mean intrinsic values
$(B-V)_\circ$=$+$0.23($\pm$0.06) at the time of $V$-band maximum, and
$(B-V)_\circ$=$-$0.02($\pm$0.04) at $t_{2}^{V}$.  The reddening we computed
for the program novae at these two epochs are in good mutual agreement. 
Because of the peculiar spectral energy distribution of novae, the intrinsic
values given by van den Bergh and Younger {\em cannot} be ported to other
color combinations using transformations calibrated on normal stars, as
discussed in Munari (2014).

The distance to a nova is usually estimated via calibrated relations (called
MMRD) between absolute magnitude at maximum and rate of decline $t_{n}^{\lambda}$,
either in the form
\begin{displaymath}
M_{\rm max}^{\lambda}\,=\,\alpha_n\,\log\, t_{n}^{\lambda} \, + \, \beta_n
\end{displaymath} 
or the stretched S-shaped curve 
\begin{displaymath}
M_{\rm max}^{\lambda}\,=\,\gamma_n\, - \delta_n\, \arctan \frac{\epsilon_n - \log\,
t_{n}^{\lambda}}{\zeta_n}
\end{displaymath} 
first introduced by Capaccioli et al.  (1989).  In computing the distances
to the program novae, we have adopted the latest available calibration by
Downes \& Duerbeck (2000) for MMRD as function of $t_{2}^{V}$ and
$t_{3}^{V}$, as well as the S-shaped curve.  These three values for the
absolute magnitude are in good mutual agreement (mean deviation from mean
value is 0.15 mag), with perhaps a slight tendency to be fainter for values
computed from $t_{2}^{V}$.  The distance given in Table~3 is computed from
the mean value of the absolute magnitude as provided by the three
$t_{2}^{V}$, $t_{3}^{V}$ and S-curve methods.

Buscombe \& de Vaucouleurs (1955) noted how the absolute magnitude 15 days
after optical maximum is similar for novae of all speed classes.  We adopt
for this the value $M_{15}^{V}$=$-$6.05 calibrated by Downes \& Duerbeck
(2000).  On average the distance computed in Table~3 from the brightness at
15 days is similar to that provided by $t_{2}^{V}$, $t_{3}^{V}$ and S-curve
methods.

In estimating the distances, the correction for extinction is computed from
the derived $E_{B-V}$ reddening and the standard $R_V$=3.1 law by Fitzpatrick
(1999), following the expression 
\begin{equation}
A(V)=3.26\times E_{B-V} + 0.033\times E_{B-V}^{2}
\end{equation}
computed by Fiorucci and Munari (2003) for the energy distribution of a nova
at the time of maximum brightness.  Compared to the total extinction along
the line of sight provided by the 3D maps of Schlegel, Finkbeiner \& Davis
(1998, hereafter SFD98) and Schlafly \& Finkbeiner (2011, SF11), the
extinction computed from Eq.(1) is nearly identical for four program novae,
and a fraction of it for other two.  The remaining program nova (V2949 Oph)
has no $B$-band photometry useful to the computation of $E_{B-V}$.

\subsection{V1534 Sco}

V1534 Sco (= Nova Sco 2014 = TCP J17154683-3128303) was discovered at
(unfiltered) 10.1 mag on 2014 Mar 26.85 by K.  Nishiyama and F. 
Kabashima (CBET 3841).  Spectroscopic classification as an He/N nova was
obtained on Mar 27.8 by Ayani \& Maeno (2014), reporting FWHM=7000 km/s
for H$\alpha$ (see also Jelinek et al.  2014).

Joschi et al. (2015) discuss the result from near-IR spectroscopy covering
the first 19 days of the outburst, following on the preliminary report by
Joschi et al.  (2014).  The near-IR spectra confirm the He/N
classification, and show emission lines characterized by a rectangular shape
with FWZI$\sim$9500 km/s and a narrow component on the top.  The positional
coincidence with a bright 2MASS cool source (J=11.255, H=10.049, Ks=9.578)
and the presence of first overtone absorption bands of CO at 2.29 microns
(as seen in M giants) led Joschi et al.  (2014) to suggest that V1534 Sco is
a nova originating from a symbiotic binary system, similar to V407 Cyg, RS
Oph and V745 Sco.

X-ray emission from the nova was observed by the Swift satellite within a few
hours of optical discovery (Kuulkers et al.  2014), corresponding to an
absorbed optically thin emission of kT = 6.4 +3.8/-2.1 keV and N$_H$ = 5.8
(+1.2/-1.0) 10$^{22}$ cm$^{-2}$ (with most of the absorption intrinsic to
the source).  X-ray emission was also recorded on the following days (Page,
Osborne and Kuulkers 2014), with the softer counts increasing as result of
decreasing absorption column, a behavior consistent with that expected for
a shock emerging from the wind of the secondary star, as expected in a nova
erupting within a symbiotic system.

The presence of a cool giant in nova V1534 Sco has however to face some
inconsistencies: ($a$) Joschi et al.  (2015) sequence of near-IR spectra
show emission lines of constant width, not the rapid shrinking associated
with the deceleration of the ejecta expanding through the pre-existing wind
of the cool giant companion, as observed in the template V407 Cyg case
(Munari et al.  2011); ($b$) the narrow peak observed by Joschi et al. 
(2015) to sit on top of the broad emission lines, and taken to represent the
flash-ionized wind of the cool giant, does not quickly disappear as
consequence of rapid recombination driven by the high electronic density, as
instead observed in other novae erupting within symbiotic binaries; and
($c$) by analogy with V407 Cyg and V745 Sco, $\gamma$-ray emission would
have been expected to arise from high velocity ejecta slamming onto the wind
of the cool companion (Ackermann et al.  2014), but no $\gamma$-ray
detection of V1534 Sco has been reported to date.

    \begin{figure}
    \includegraphics[width=84mm]{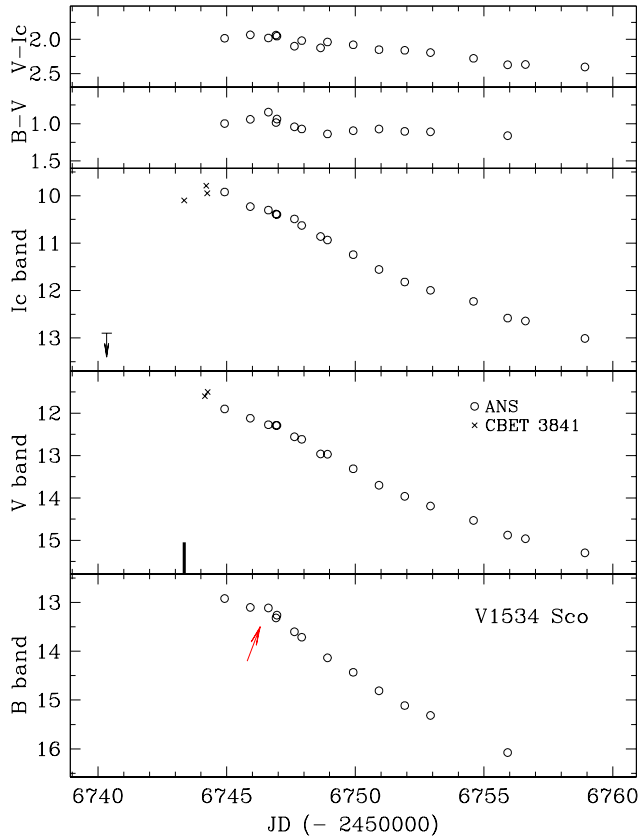}
    \caption{$B\,VI$ photometric evolution of V1534 Sco (= Nova
    Sco 2014 = TCP J17154683-3128303). Data from CBET 3841 are reported as
    crosses and upper limit. The solid vertical line in the $V$-band panel
    marks the time of nova discovery. The arrow points to the hiccup
    considered in Figure~3.}
    \end{figure}

\subsubsection{The lightcurve}

The lightcurve of V1534 Sco is presented in Figure~2 and the basic
parameters extract from it are listed in Table~3.  Our photometric
monitoring commenced within one day of the announcement of its discovery. 
Our observations continued indeed for a longer period than shown in the
figure, but we refrain from plotting or tabulating such noisy late data,
which are best described as a rapid flattening of the lightcurve toward the
asymptotic values $B$$\sim$19.6, $V$$\sim$18.3, and $I$$\sim$14.3.  This
flattening is artificial and can be ascribed to (1) the stable contribution
from the 2MASS cool source which dominates in the $I$ band, and (2) the
unresolved contribution in $B$ and $V$ bands by several unrelated field
stars which lie within $\sim$4 arcsec of the nova.  The crowding is so
severe in the immediate surrounding of the nova that attempts to disentangle
it via PSF-fitting proved unconverging on our images.

The time and brightness of maximum in $I$ band is well constrained in
Figure~2, and by similarity we assume the earliest two $V$ points plotted in
Figure~2 to mark the actual maximum in the $V$ band.  Lacking $B$-band data
for the maximum, the $E_{B-V}$ reddening can be estimated only from $B-V$
color at $t_{2}^{V}$, providing $A_V$=3.66 from Eq.(1).  From this and the
extremely fast decline times listed in Table~3, the distance to this nova
would turn unconfortably large, $\geq$30 kpc, placing it far beyond the
Galactic Bulge against which it is seen projected, at a hight of $\geq$2 kpc
above the Galactic plane.  This is clearly an unlikely location.

    \begin{figure}
    \includegraphics[width=84mm]{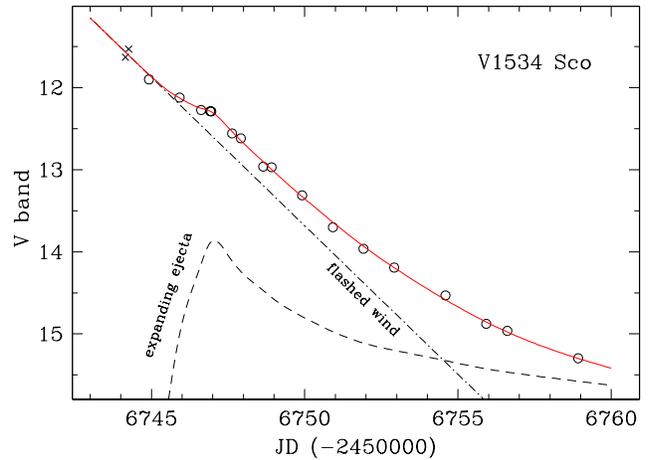}
    \caption{Fit to the $V$-band lightcurve of V1534 Sco (from Figure~2),
    deconvolved into the lightcurve of the expanding ejecta and that of
    the recombining wind of the cool giant (for an $e$-folding time of 3
    dyas) ionized by the initial flash of the nova (see sect. 4.2.1 for
    details).} 
    \end{figure}

The $B-V$ color measured for this nova at $t_{2}^{V}$ seems strongly
influenced by the presence of the cool giant, the severe crowding and the
contribution by recombining flash-ionized wind of the companion, to the
point of fooling the comparison with van den Bergh and Younger (1987)
intrinsic colors.  For similar reasons the brightness at 15 days ($V$=16.35
mag) appears useless in estimating the distance.  If for the extinction we
adopt instead the values given by SFD98 and SF11 and reported in Table~3,
the distance to the nova results in 5.4 and 9.4 kpc, respectively. 
In Table~3 we list the average 7.4 kpc value, well compatible with a
partnership to the Galactic Bulge.

It is worth noting that the lightcurve of V1534 Sco displays a distinctive
hiccup marked by the red arrow in Figure~2.  Its strength is
wavelength-dependent, descreasing from $B$ to $I$ band.  For sake of
discussion, we have fitted the $V$-band lightcurve of V1534 Sco 
with the combination of two sources: the flash-ionized wind of the cool
giant and the expanding nova ejecta.  The latter is obtained as the
difference (computed in the flux space) between the observed lightcurve and
the exponential decline from the flash-ionized wind.  In Figure~3 we present
the results for a recombination $e$-folding time of 3 days, corresponding to
an electron density of 1.5$\times$10$^7$ cm$^{-3}$ at the peak of the
ionization and assuming an electron temperature of 10\,000 K.  The fit looks
excellent, but this is hardly a proof of its uniqueness.  Given the fact
that the near-IR observations by Joschi et al.  (2015) did not detected a
decelleration of the ejecta, it makes sense to treat the lightcurve derived
in Figure~3 for the expanding ejecta as that of the nova proper.  In this
case the $V$ band maximum was reached on JD=2456747.0 at magnitude
$\sim$13.8, with a decline rate $t_{2}^{V}$$\sim$13 days.  Adopting the
larger extinction from SFD98, the distance ($\sim$10 kpc) would still be
compatible with a partnership to the Bulge (the fainter apparent magnitude
is compensated for by a similarly fainter absolute value implied by the
slower decline rate).

\subsection{V1535 Sco}

    \begin{figure}
    \includegraphics[width=84mm]{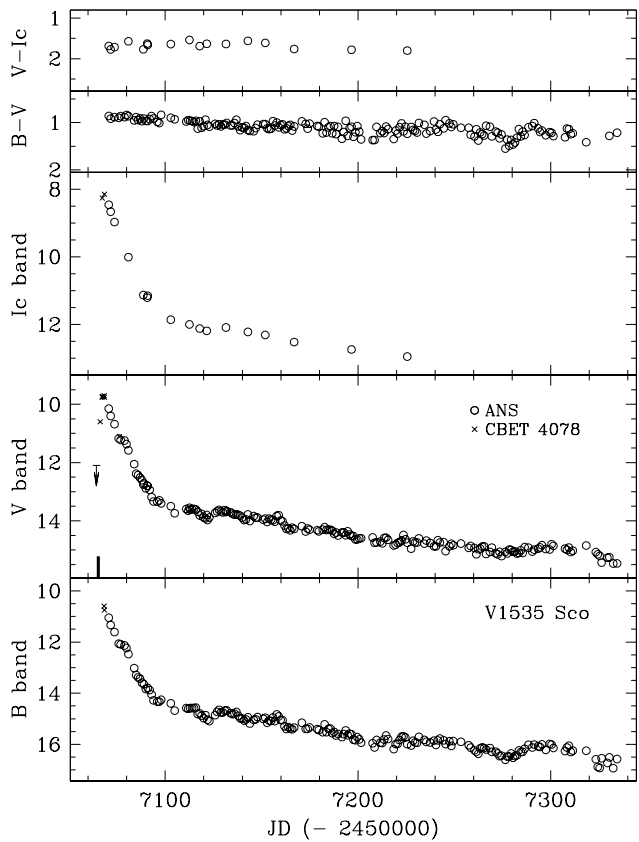}
    \caption{$B\,VI$ photometric evolution of V1535 Sco (= Nova
    Sco 2015 = PNV J17032620-3504140).  The solid vertical line in the
    $V$-band panel marks the time of nova discovery.}
    \end{figure}

  V1535 Sco (= Nova Sco 2015 = PNV J17032620-3504140) was discovered by T. 
Kojina on 2015 Feb 11.837 (CBET 4078) and soon confirmed spectroscopically
by Walter (2015) as an He/N nova.  Nelson et al.  (2015) performed X-ray and
radio observations within a few days of discovery, and found the initial
presence of hard, absorbed X-rays and synchrotron radio emission that
suggested the nova erupted in a symbiotic binary, with collision between the
ejecta and the cool giant wind shock-heating plasma and accelerating
particles.  The suggestion about the presence of a cool giant companion was
made also by Walter (2015).  The synchrotron radio component rapidly
declined during the following days while more conventional thermal free-free
emission emerged (Lindford et al.  2015).  Near-IR spectral monitoring by
Srivastava et al.  (2015a,b) revealed a progressive narrowing of the
never-too-broad emission lines from FWHM$\sim$2000 down to 500 km/s,
indicating a decelerating shock as the nova ejecta collide with and are
slowed down by the wind of the giant companion.  The extra brightness of the
progenitor in quiescence H$\alpha$ Super-COSMOS images was taken by
Srivastava et al.  (2015b) as a further evidence of a symbiotic nature. 
Linear polarization measurements in BVRI bands at seven consecutive dates in
February were reported by Muneer, Anupama and Raveendran (2015) who
concluded that, even if not corrected for interstellar polarization, the
data support intrinsic polarization.

    \begin{figure}
    \includegraphics[width=84mm]{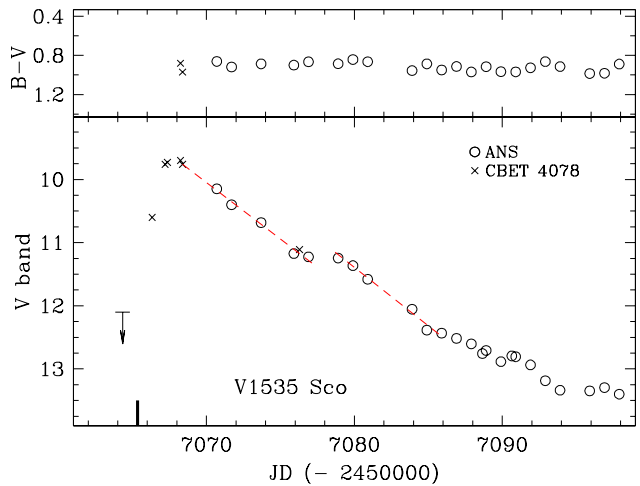}
    \caption{A zoomed portion around maximum of the $V$-band lightcurve of
    V1535 Sco (= Nova Sco 2015 = PNV J17032620-3504140) from Figure~4.  The
    solid vertical line marks the time of nova discovery, and the upper
    limit is from CBET 4078 ported to $V$ band.  The dashed lines aim to
    highlight the plateau around JD=2457078, which is discussed in sect
    4.3.1.}
    \end{figure}

\subsubsection{The lightcurve}

The lightcurve resulting from  our year-long monitoring of V1535 Sco is
presented in  Figure~4, and the basic nova parameters are summarized in 
Table~3.

The overall lightcurve of V1535 Sco looks pretty standard, with a faster
decline during the initially optically thick conditions followed by a
slower descent during the later optically thin phase.  The transition from
optically thick to thin ejecta occoured 27 days past and
$\Delta$mag=3.60 below the $V$-band maximum.  Such a well behaving
transition is usually seen in FeII novae (eg.  McLaughlin 1960), but much
less frequently in He/N novae.  The latter usually expel less material at
larger velocity and higher ionization compared to FeII counterparts, and
their ejecta reach optically thin conditions much closer to maximum
brightness.

A noteworthy feature of the lightcurve is the temporary dip that the nova
went through around April 8 (JD=2457121), which developed at constant colors
during the optically thin phase, as if for some time the ejecta were exposed
to a lower flux of ionizing photons from the central source.  A second,
wider, and stronger dip which occoured in mid-September (around JD=2457280),
was instead strongly color-dependent.  Their interpretation would
require access to a detailed spectroscopic monitoring that we lack. 
Overall, the lightcurve of V1535 Sco during the optically thin phase has
been "bumpy", well beyond the measurement errors.

Figure~5 zooms on the early portion of the $V$-band lightcurve, to highlight
the plateau lasting a couple of days around Feb 24, or ten days past and
1.5 mag below optical maximum. Two possible interpretations come to mind,
but both have their shares  of problems. 

First, the plateau could represent the same type of transition discussed in
Figure~3 above for V1534 Sco, namely the emission from expanding ejecta
overtaking that of flash-ionized wind of the cool companion.  This contrasts
with the long delay past maximum, requiring a rising time to maximum for the
ejecta ($\geq$10 days) which is more typical of FeII events than He/N for
which it is generally an order of magnitude faster.  This could be
counter-argumented by noting that ($i$) the initial He/N spectral
classification for V1534 Sco could have been fooled by the dominating
emission from the flash-ionized wind, and ($ii$) the narrowness of the
emission lines, their Gaussian-like shapes and the presence of P-Cyg
absorptions observed in the near-IR by Srivastava et al.  (2015b) are more
typical of FeII novae, while He/H tend to show much broader and rectangular
emission lines with no P-Cyg absorption components (Banerjee \& Ashok 2012). 
It will be interesting to carefully inspect, when they will be eventually
published, optical spectra taken over a protracted interval of time to
ponder the spectral classification of the expanding nova ejecta separately
from that of the flash-ionized wind.

    \begin{figure}
    \includegraphics[width=84mm]{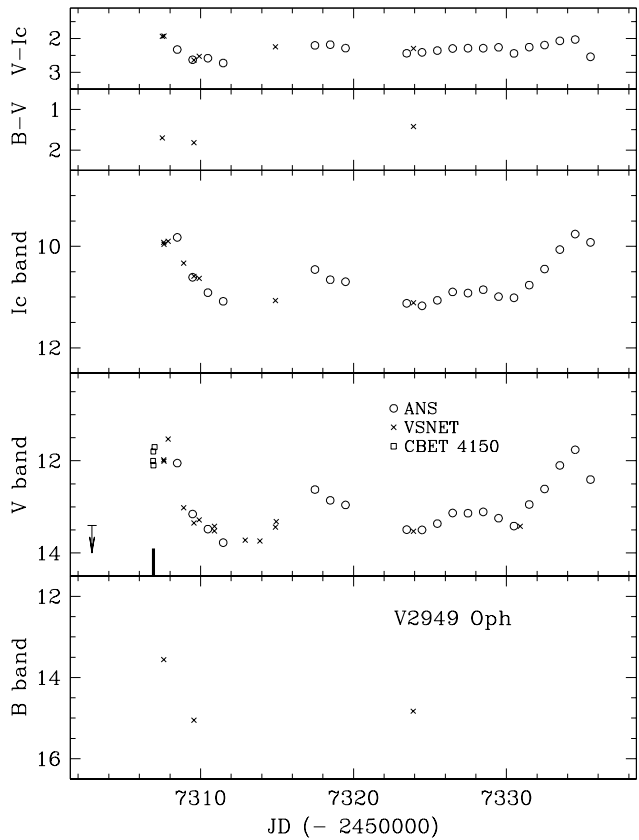}
    \caption{$B\,VI$ photometric evolution of V2949 Oph (= Nova Oph 2015 N.2
    = TCP J17344775-2409042).  Data from CBET 4150 are reported as open
    squares and upper limit.  The solid vertical line in the $V$-band panel
    marks the time of nova discovery.}
    \end{figure}

Secondly, a similar plateau has been sometimes observed in novae during the
super-soft phase, when optically thin ejecta are exposed to the hard
radiation field of the central white dwarf still burning nuclearly at its
surface.  The consequent input of ionizing photons spreading through the
ejecta counter-balances the recombination of ions.  The plateau is usually
terminated by either rapid dilution in fast expanding and low mass ejecta
(as observed during the 2016 outburst of the recurrent nova LMC 1968, Munari et
al.  2016a), or by switching off the nuclear burning on the WD (as in U Sco,
Osborne et al.  2010).  The problem in this case is that the plateau
occoured two weeks before the ejecta turned optically thin on day 27 past
optical maximum.  A way out could be a highly structured, non-spherical
shape of the ejecta, with optical thickness strongly dependent on angular
coordinates.  Hints in favor of such an arrangement are the fact that the nova
erupted within the pre-existing wind of the giant companion, and optical
(Walter 2015) as well as near-IR (Srivastava et al.  2015b) spectra which 
present weak emission components separated from the corresponding main ones.
 
The reddening estimated from nova colors and the total extinction along the
line of sight deduced from SFD98 and SF11 maps are in excellent agreement
(Table~3), and the derived distance places V1535 Sco at the distance of the
Galactic Bulge against which it is seen projected.

\subsection{V2949 Oph}

V2949 Oph (= TCP J17344775-2409042 = Nova Oph 2015 N.2) was discovered on
2015 Oct 11.41 by K.  Nishiyama and F.  Kabashima (CBET 4150), and
confirmed spectroscopically on Oct 12.42 by Ayani (2015).  Low expansion
velocity, heavy reddening and a Fe-II spectral class was reported by
Campbell et al.  (2015) from Oct 11 spectroscopic observations, while
Littlefield and Garnavich (2015) from Oct 11.99 observations estimated in
900 km/s the FWHM of H$\alpha$ emission and $-$800 km/s the velocity of its
P-Cyg absorption component.

\subsubsection{The lightcurve}

Our lightcurve for V2949 Oph in presented in Figure~6.  We begun the
observations as night settled on Oct 12.98, soon after spectroscopic
confirmation was circulated, and continued them until Nov 9, when Solar
conjunction prevented further data to be collected.  This is the only
program nova that was not observed also in $B$ band.

The lightcurve shows the nova fluctuating by $\Delta V$$\sim$2 mag around
maximum brightness.  Similar peak brightness was reached on Oct 12.38 at
$V$=11.41 and on Nov 7.99 at $V$=11.76.  The first has been taken - somewhat
arbitrary - as the true maximum, so that the brightness 15 days past it can
be used to estimate a distance of 8.4 kpc (cf.  Table~3), which places the
nova right at the distance of the Galactic center.  The reddening resulting
from $B-V$ color around maximum (cf.  Figure~6) indicates an extinction
$A_V$$\sim$4.9, uncomfortably in excess of the total value along the line of
sight $A_V$$\sim$3.39 from SFD98 and $A_V$$\sim$2.88 from SF11 maps. 
Because the very few $B$ magnitudes used in this exercise are not ours and
come instead from VSNET observers (who did not provide details on their data
reduction procedures and adopted comparison sequence), we will make no
further use of these $B$-band data.

\subsection{V3661 Oph}

    \begin{figure}
    \includegraphics[width=84mm]{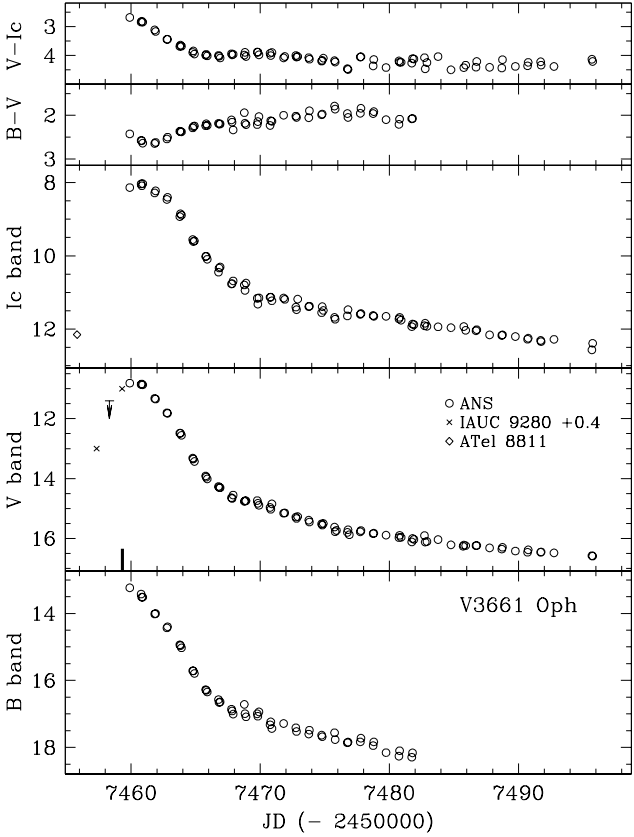}
    \caption{$B\,VI$ photometric evolution of V3661 Oph (= Nova Oph 2016 =
    PNV J17355050-2934240).  The data from IAUC 9280 (including the upper
    limit) refer to unfiltered observations, which have been shifted by 0.4
    mag to fit the $V$-band lightcurve.  The solid vertical line in the
    $V$-band panel marks the time of nova discovery.}
    \end{figure}

V3661 Oph (= PNV J17355050-2934240 = Nova Oph 2016) was discovered in
outburst by H.  Yamaoka on Mar 11.81 (CBET 4265).  A preliminary
spectroscopic classification as a nova was derived by Munari et al. 
(2016b) from a very low S/N spectrum, with later IR and optical spectra by
Srivastava et al.  (2016) and Frank et al.  (2016) fixing the spectral class
to FeII.  All three spectral sources concur on a highly reddened continuum,
FWHM$\sim$1000/1400 km/s for Balmer emission lines and a velocity separation
of $\sim$950 km/s between the emission and absorption components of the
P-Cyg profile affecting most of the lines.  A pre-discovery OGLE-IV
observation at $I$=12.15 on March 8.31 was reported by Mroz \& Udalski
(2016a) who noted the absence of the progenitor in OGLE deep template
images, meaning it was fainter than 22 mag in $I$ band.  A pre-discovery
observation by ASAS-SN of the nova on March 10.85 has been noted by
Chomiuk et al.  (2016).  Finally, Muneer \& Anupama (2016) reported
significant linear polarization in $V$$R$$I$ photometric observations of
V3661 Oph obtained from March 13 to 19, that they interpret as arising
primarily in the interstellar medium given the high reddening suffered by
the nova.

    \begin{figure}
    \includegraphics[width=84mm]{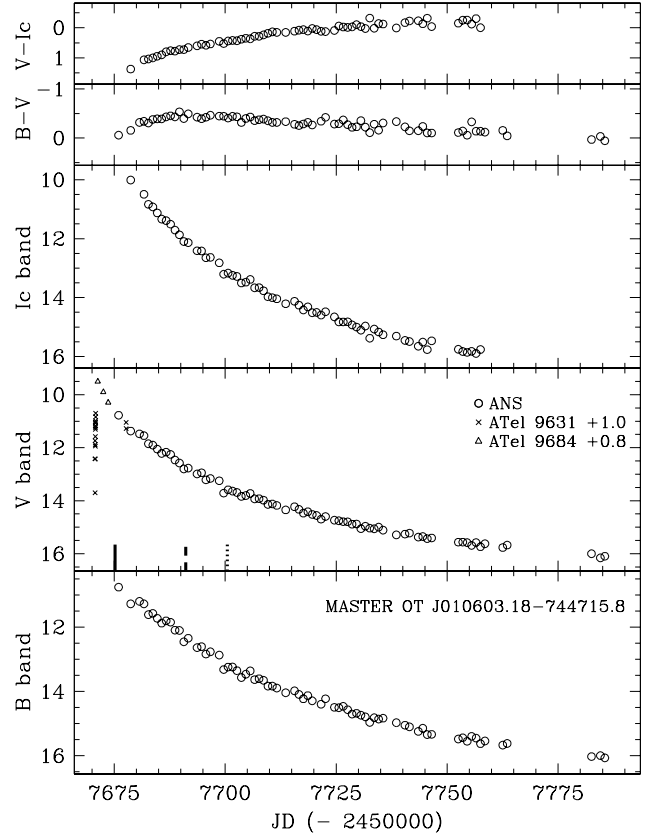}
    \caption{$B\,VI$ evolution of MASTER OT J010603.18-744715.8 (= Nova SMC
    2016).  The solid, dashed and dotted vertical lines in the $V$-band
    panel marks respectively: announcement of the nova by Shumkov et al. 
    (2016), appearance of nebular emission lines in optical spectra
    according to Williams \& Darnley (2016b), and emergence of super-soft
    X-ray emission following Page at al.  (2016).  Pre-discovery unfiltered
    data from Lipunov et al.  (2016) and Jablonski \& Oliveira (2016) are
    plotted on the $V$-band lightcurve after applying the offsets indicated
    (in magnitudes, see sect 4.6.1 for details).}
    \end{figure}

\subsubsection{The lightcurve}

Our lightcurve for V3661 Oph is presented in  Figure~7, and the basic nova
parameters are summarized in Table~3 as for the other program objects.

The lightcurve looks particularly well behaving, almost a textbook example
for a FeII nova. The clear dependence on wavelength of the time of maximum
brightness will be discussed in sect. 5 below, in parallel with the similar
case for TCP J18102829-2729590. The transition from optically thick to thin 
ejecta occoured 6.0 days past and $\Delta$mag=3.35 below $V$-band maximum.

With $t_{2}^{V}$=3.9 and $t_{3}^{V}$=5.7 days, V3661 Oph is probably the
fastest known nova of the FeII type, and a very fast one even compared with
He/N recurrent novae like U Sco.  It is by far the nova with the reddest
colors and therefore the highest extinction among the program ones, with a
mean observed color $B-I$$\sim$6.25, as averaged along the whole lightcurve. 
SFD98 and SF11 maps also suggest an extremely large total extinction along
the line of sight to V3661 Oph.  Finally, the short distance derived for
this nova places it much closer than the Bulge and within the Galactic disk.

\subsection{MASTER OT J010603.18-744715.8}

MASTER OT J010603.18-744715.8 was discovered (at unfiltered 10.9 mag) on
2016 Oct 14.19 by the MASTER-OAFA autodetection system and announced by
Shumkov et al (2016) on Oct 14.34.  ANS Collaboration monitoring begun on
Oct. 14.51. Detection of the progenitor at mean $I$=20.84 and
($V$$-$$I$)=+0.16 on archive OGLE-IV observations was reported by Mroz and
Udalski (2016b), with hints of semi-regular variability of a timescale
of 20-30 days.

    \begin{figure}
    \includegraphics[width=84mm]{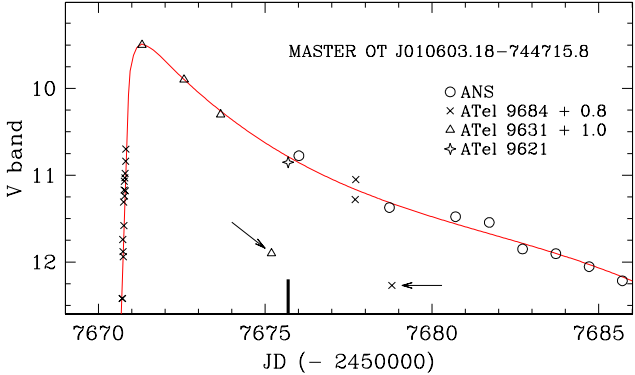}
    \caption{A zoomed portion around maximum of the $V$-band lightcurve of
    MASTER OT J010603.18-744715.8 from Figure~8.  The
    solid vertical line marks the time of nova discovery. The continous line
    is a polynomial fit to the data to guide the eye. The arrows point to
    discordant data from ATel 9631 and ATel 9684 discussed in the text.}
    \end{figure}

Lipunov et al. (2016) found pre-discovery MASTER images that show how the
nova was already declining from maximum when first noticed.  An image for Oct
9.81 recorded the nova at (unfiltered) 8.5 mag, declining to 8.9 mag on
Oct 11.07 and 9.3 mag on Oct 12.16.  Robotic DSLR-camera monitoring of the 
SMC was inspected by Jablonski \& Oliveira (2016) to obtain the (unfiltered)
brightness profile of the rise toward maximum of the nova during Oct 9.  The
nova was fainter than 13.2 mag on Oct 9.197, first detected at 12.9 mag on
Oct 9.210, and last measured at 9.90 mag on Oct 9.325.

    \begin{figure*}
    \includegraphics[width=170mm]{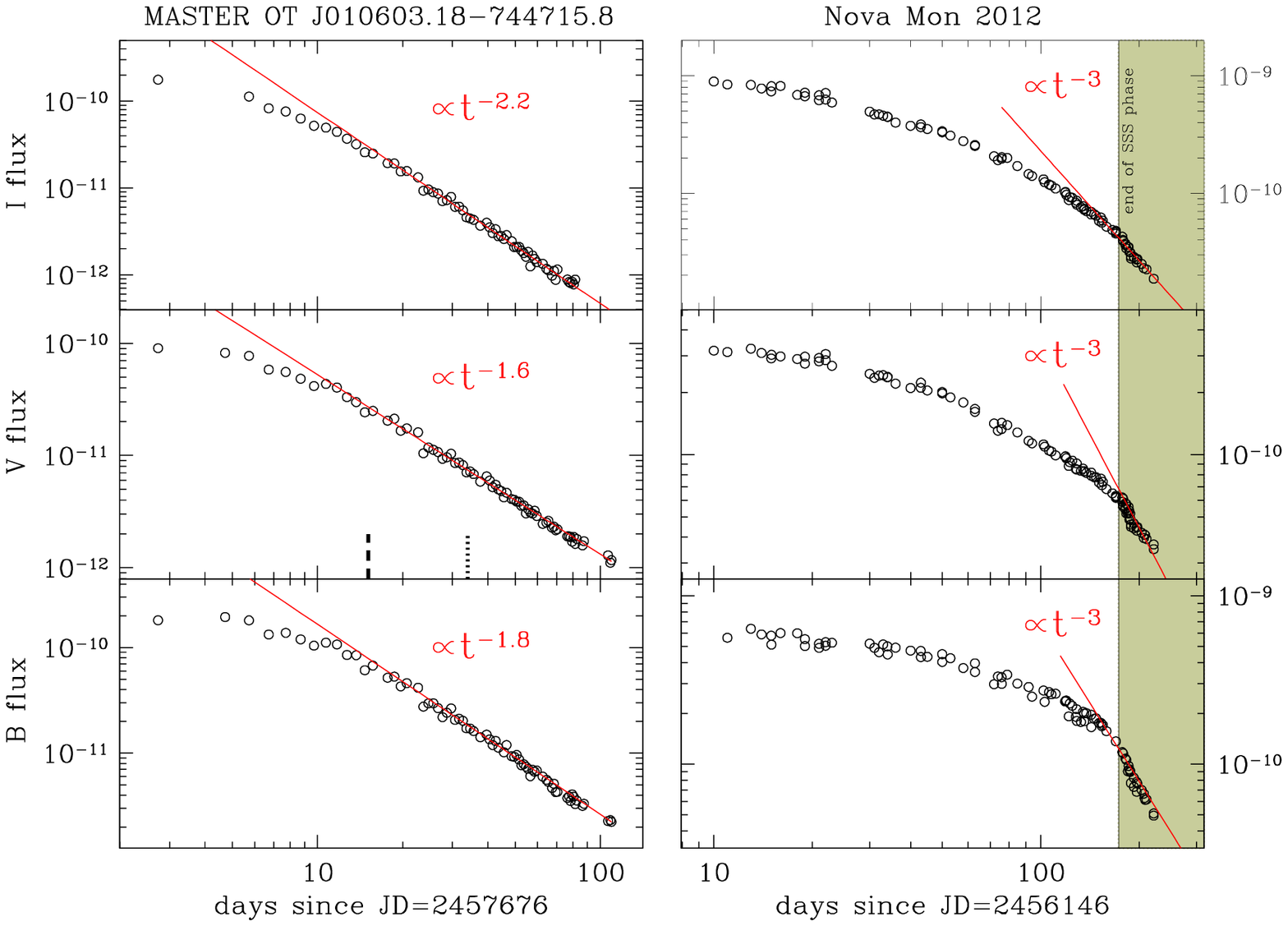}
    \caption{{\em Left:} evolution of flux through $B\,VI$ bands for MASTER
    OT J010603.18-744715.8 (= Nova SMC 2016).  The lines are linear fits to
    flux at later epochs, providing the indicated decline rates.  The dashed
    and dotted vertical lines in the $V$-band panel marks respectively the
    appearance of nebular emission lines in optical spectra (Williams and
    Darnley 2016b), and emergence of super-soft X-ray emission (Page at al. 
    2016).  {\em Right:} the same for Nova Mon 2012 (data from Munari et al. 
    2013a) to highlight the change in the slope following the end of the
    super-soft phase and thus the end of nuclear burning on the central
    white dwarf (shadowed region to the right).} 
    \end{figure*}

Spectroscopic confirmation was obtained by Williams \& Darnley (2016a) on
Oct 14.70.  They measured FWHM$\sim$3700 km/s for Balmer lines and
classified the nova type as He/N.  Following their description of the
observed emission lines, the signatures in favor of a He/N class appears
however weaker than typical for this type, with some room left for a FeII
classification.  Darnley \& Williams (2016b) reported on their continued
spectroscopic monitoring of the nova till Oct 29, noting the disappearance
of P-Cyg absorptions and the emergence of HeI 5876, 7065 and of $[$OIII$]$
4959/5007, from which they infer the nova had entered the nebular phase.  It
should be noted that the presence of HeI emission lines at the time $[$OIII$]$
emerges is standard for FeII novae, and that presence of P-Cyg absorption
and $[$OIII$]$ nebular lines are more typical of FeII than He/N novae
(Williams 1992).  In addition, the FWHM$\sim$3700 km/s observed for
Balmer lines is close to the low limit for typical He/N novae while still well
suited for FeII ones.

Nova MASTER OT J010603.18-744715.8 has been intensively monitored in X-rays.
Early Swift observations on Oct 15 failed to detect X-ray emission (Kuin et
al. 2016). Rapidly brightening soft X-ray emission was detected by Swift 
starting with Nov 7 (Page et al. 2016). Chandra observations for Nov
17-18 (Orio et al. 2016a) confirmed the super-soft bright emission, which
continued well into the Chandra observation for 2017 Jan 4 (Orio et al.
2016b), when a preliminary fit to the spectra supports an increase from
650,000 to 750,000 K for the temperature of the white dwarf.

\subsubsection{The lightcurve}

The lightcurve of MASTER OT J010603.18-744715.8 is particoularly simple and
smooth, and it is presented in Figure~8, with the basic parameters extracted
from it summarized in Table~3.  Figure~9 zooms on the phase of maximum,
which was very brief with an extremely fast rise toward it, as pre-discovery
observations by Lipunov et al.  (2016) and Jablonski \& Oliveira (2016) help
to constrain.  These unfiltered observations (i.e.  white light, and
therefore strongly skewed toward red wavelengths where CCD sensitivity
peaks) require a large color correction to be properly plotted on the
$V$-band plane, because of the remarkably blue colors for the nova resulting
from the very low reddening toward SMC.  The color corrections are given in
Figure~9, and have been derived by continuity in comparison with our
properly calibrated photometry.  The latest observation listed by both
Lipunov et al.  (2016) and Jablonski \& Oliveira (2016) are clearly off the
otherwise well behaving lightcurve of the nova (the two arrows in Figure~9
point at them), and are ignored as erroneous data.

\subsubsection{Super-soft X-rays and rate of decline}

Our $B\,VI$ data are transformed into absolute fluxes (erg cm$^{-2}$
s$^{-1}$) and log-log plotted against time on the left panel of Figure~10. 
For a comparison, the same is done on the right panel for Nova Mon 2012
(data from Munari et al.  2013a).  The shaded area in the figure marks the
time after the super-soft X-ray emission had ceased (Nelson et al.  2012,
Page et al.  2013).  The phase of super-soft X-ray emission corresponds to
optically thin ejecta permeated by the hard radiation from the central white
dwarf undergoing stable nuclear burning at its surface (Krautter 2008,
Schwarz et al.  2011).  Such ionizing radiation partially counter-balance
the recombination in the expanding ejecta, keeping high their emissivity and
flattening the decline rates.  When, with the end of the super-soft phase
this hard radiation input ends, the emissivity of the ejecta rapidly settles
onto the pure recombination rate $\propto t ^{-3}$, which is precisely what
Nova Mon 2012 duly did.  For MASTER OT J010603.18-744715.8, as soon as it
entered the nebular phase and super-soft X-ray emission emerged (Williams
and Darnley 2016b, Page et al.  2016), the decline in flux rapidly settled
on a rate kept stable for all the period covered by our observations:
$\propto t ^{-1.8}$, $\propto t ^{-1.6}$ and $\propto t ^{-2.2}$, for $B$,
$V$ and $I$ respectively.  The rates are slightly different from band to
band, depending on the fractional contribution of continuum and emission
lines, which decline at different speed as the degree of ionization and
electron density change through the ejecta.  The fact that these rates are
much flatter than $\propto t ^{-3}$ is interpreted as an evidence that
nuclear burning was still up and running on the surface of the central WD at
the time of our last observations.  When the nuclear burning will eventually
end, it is expected that the decline in brightness of MASTER OT
J010603.18-744715.8 will accellerate to $\propto t ^{-3}$, as seen in Nova
Mon 2012.

\subsection{TCP J18102829-2729590}

TCP J18102829-2729590 was discovered on 2016 October 20.383 at 10.7 mag by
K.  Itagaki (cf.  CBET 4332).  Mroz et al.  (206) derived astrometric
coordinates from OGLE-IV $I$-band images as RA=18:10:28.29 and
DEC=$-$27:29:59.3, and noted that the progenitor is undetected in
pre-outburst OGLE deep template images, meaning $I$$>$22 mag.  Spectroscopic
classification as a FeII-class nova was obtained by Lukas (2016). 
$\gamma$-ray emission from this nova has been detected by Fermi-LAT (Li \&
Chomiuk 2016).

\subsubsection{The lightcurve}

Our daily-mapped lightcurve of TCP J18102829-2729590 is shown in Figure~11. 
It extends over a whole month and fully covers the phase of maximum
brightness and decline well past $t_{3}^{V}$.  It is real smooth and
characterized by a rapid initial rise and two distinct maxima.  Our
monitoring was stopped by Solar conjunction when the nova was still bright. 
As for the other novae, the parameters extracted from the lightcurve are
listed in Table~3.  The distances derived from $t_{2}^{V}$, $t_{3}^{V}$ are
$V_{15}$ are dependent from which of the two maxima is taken as reference. 
The average of the values listed in Table~3 is 8.1 kpc, right that of the
Bulge against which the nova is seen projected.  The partnership to the
Bulge is confirmed by the photometric reddening of the nova that equals the
total extinction along the line of sight from the 3D maps of SFD98 and SF11.

There is a striking difference between the two maxima displayed by TCP
J18102829-2729590: the first one is markedly wavelength dependent, the other
is not.

The wavelength dependence of the first maximum manifests in a time-delay of
$\sim$1 day between peak brightness in $B$ and $I$ bands, as noted above for
V3661 Oph.  As discussed in sect.  5 below, this is a characteristic of the
initial fireball expansion of the ejecta, with maximum representing the time
of largest angular extension for the pseudo-photosphere that is optically
thick at the given wavelength.  The independence from wavelength of the
second maximum suggests it is of a different physical nature, which will be
discussed in sect.  6 below.

    \begin{figure}
    \includegraphics[width=84mm]{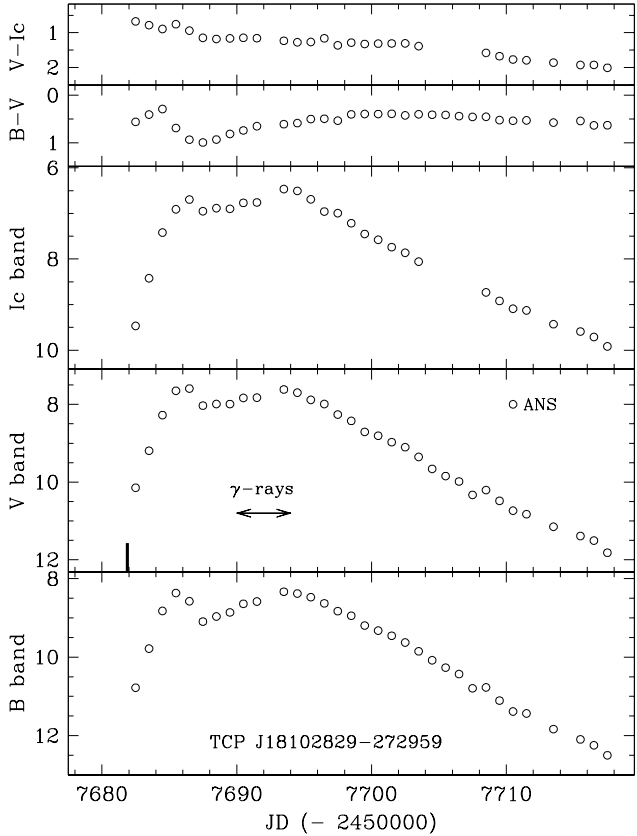}
    \caption{$B\,VI$ photometric evolution of nova TCP J18102829-2729590. 
    The solid vertical line in the $V$-band panel marks the time of nova
    discovery.  The left-right arrow highlights the time interval during which
    $\gamma$-ray emission from the nova was detected by Fermi-LAT (Li \& 
    Chomiuk 2016).}
    \end{figure}

\subsection{ASASSN-16ma}

ASASSN-16ma was discovered at $V$$\sim$13.7 in ASASSN-CTIO images obtained
on 2016 Oct 25.02, brightened to $V$$\sim$11.6 a day later and was
undetected ($V$$>$17.3) on Oct 20.04 (Stanek et al.  2016).  Its coordinates
were originally given as RA=18:20:52.12 and DEC=$-$28:22:13.52, which Saito
et al.  (2016) adopted to identify the likely progenitor in the VVV Survey,
as a source possibly consisting of two unresolved components of combined
brightness $z$ = 18.8, $Y$ = 18.5, $J$ = 18.1, $H$ = 17.8, and $K_s$ = 17.6
mag.  Mroz et al.  (2016) remeasured the position of the nova on OGLE-IV
$I$-band images and derived a different astrometric position RA=18:20:52.25
and DEC=$-$28:22:12.1, which is 2.4 arcsec away from the initial ASASSN-CTIO
one.  The progenitor is not visible in pre-outburst OGLE-IV survey images,
meaning that is was fainter than $I$$>$22 mag and, therefore, the star
proposed by Saito et al.  is an unrelated field star.

A low resolution spectrum obtained on Oct 27.5 by Luckas (2016) showed the
object to be a FeII-class nova.  A month later, on Nov 23.1, Rudy et al.  
(2016) obtained an optical/near-IR spectrum of ASASSN-16ma that confirmed
the FeII classification and was characterized by prevailing low-expansion
velocity and low-excitation conditions.  $\gamma$-ray emission from
ASASSN-16ma was discovered by Li, Chomiuk \& Strader (2016) while they were
monitoring with Fermi-LAT the nearby nova TCP J18102829-2729590, described
in the section above.  ASASSN-16ma remained undetected by Fermi-LAT until
Nov 8 (JD=2457701) when it suddenly turned into a strong $\gamma$-ray
source, remaining active (although declining) for the following 9 days (Li
et al.  2016).

    \begin{figure}
    \includegraphics[width=84mm]{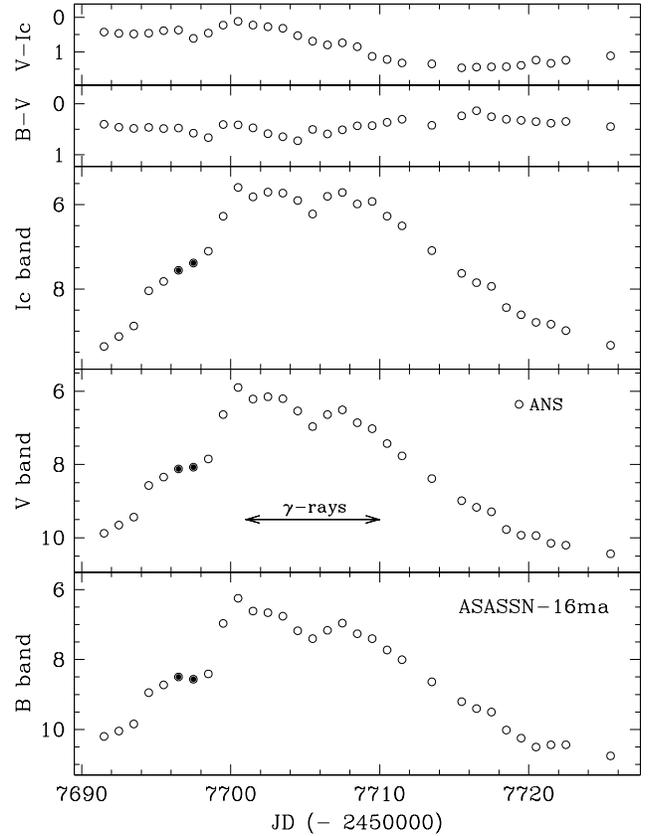}
    \caption{$B\,VI$ photometric evolution of ASASSN-16ma. The left-right
    arrow marks the time interval during which $\gamma$-ray emission from the
    nova was detected by Fermi-LAT (Li, Chomiuk \& Strader 2016, Li et al.
    2016). For the meaning of the two filled dots (the same plotted also in
    Figure 15), see text (sect. 6).}
    \end{figure}

    \begin{figure}
    \includegraphics[width=84mm]{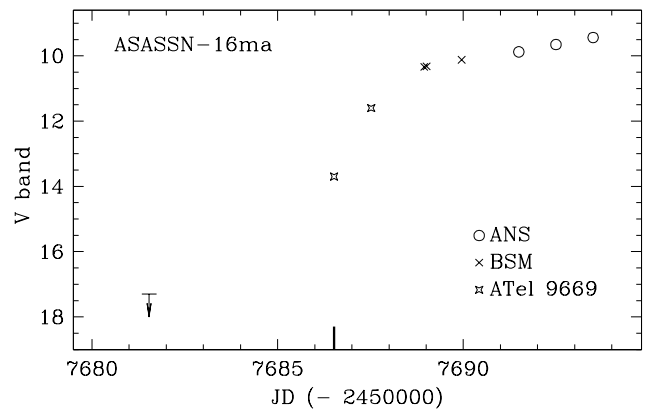}
    \caption{Zooming on the early evolution of ASASSN-16ma. The solid
    vertical line marks the time of nova discovery. BSM is the AAVSO bright
    star monitor (by A. Henden).}
    \end{figure}

\subsubsection{The lightcurve}

Our daily-mapped lightcurve of ASASSN-16ma is shown in Fig.~12. It extends
over a whole month and covers the initial rise, the phase of maximum and
decline well past $t_{3}^{V}$.  Our monitoring was stopped by Solar
conjunction when the nova was still bright.  A zoomed view of the initial
rise in brightness is given in Figure~13, where our $V$-band observations are
combined with literature data.

The lightcurve of ASASSN-16ma started as a simple one.  The two observations
for November 4.0 and 5.0 (JD=2457696.5 and 97.5) are highlighted with filled
dots in Figure~12.  They are characterized by the same dependence on
wavelength as for the maximum brightness of V3661 Oph (Figure 7) and the
first maximum of TCP J18102829-2729590 (Figure~11), namely a time delay of
$\sim$1 day between the maximum in $B$ and $I$ bands.  We believe these
filled dots trace the normal fireball maximum ASASSN-16ma initially went
through.  In support of this interpretation it is worth noticing that for
the two $\gamma$-ray program novae, both belonging to the Bulge and affected by
a similarly low reddening, the first maximum occurred at a similar
brightness: $V$=8.1 for ASASSN-16ma and $V$=7.6 for TCP J18102829-2729590.

Soon after the passage through the fireball maximum,  ASASSN-16ma raised to a
second and brighter maximum, composed by two peaks.  As for TCP
J18102829-2729590, the distance to ASASSN-16ma derived from $t_{2}^{V}$,
$t_{3}^{V}$ and $V_{15}$ depends from which of these two peaks is taken as
reference.  Choosing the first one (at JD=2457700.5) returns a distance
shorter than that of the Bulge, and a larger one selecting the second (at
JD=2457707.5).  The average is 8.3 kpc, placing also ASASSN-16ma at the
distance of the Bulge against which the nova is seen projected.  Similarly
to TCP J18102829-2729590, the partnership to the Bulge is confirmed by the
photometric reddening of ASASSN-16ma that equals the total extinction along
the line of sight from the 3D maps of SFD98 and SF11.  As for TCP
J18102829-2729590, the second maximum displayed by ASASSN-16ma is discussed
in section 6 below.

\section{The fireball expansion}

The initial photometric evolution of a nova is characterized by the rise
toward maximum, the maximum itself, and the settling onto decline.  The rise
toward maximum is rarely mapped at optical wavelengths (Seitter 1990),
because it is usually very fast (a matter of few days or even hours)
and the discovery of the nova has a higher chance to occur when the object
is at peak brightness (especially so in crowded fields).

Nonetheless, sometimes the conditions are just right to cover the transit of
a nova through optical maximum.  For the seven program novae, this is the
case for TCP J18102829-2729590 and V3661 Oph, and marginally so for others. 
Figure~14 presents a zoom on their lightcurves around optical maximum.  The
obvious feature is how the maximum brightness occurs at later times with
increasing wavelength.

    \begin{figure}
    \includegraphics[width=84mm]{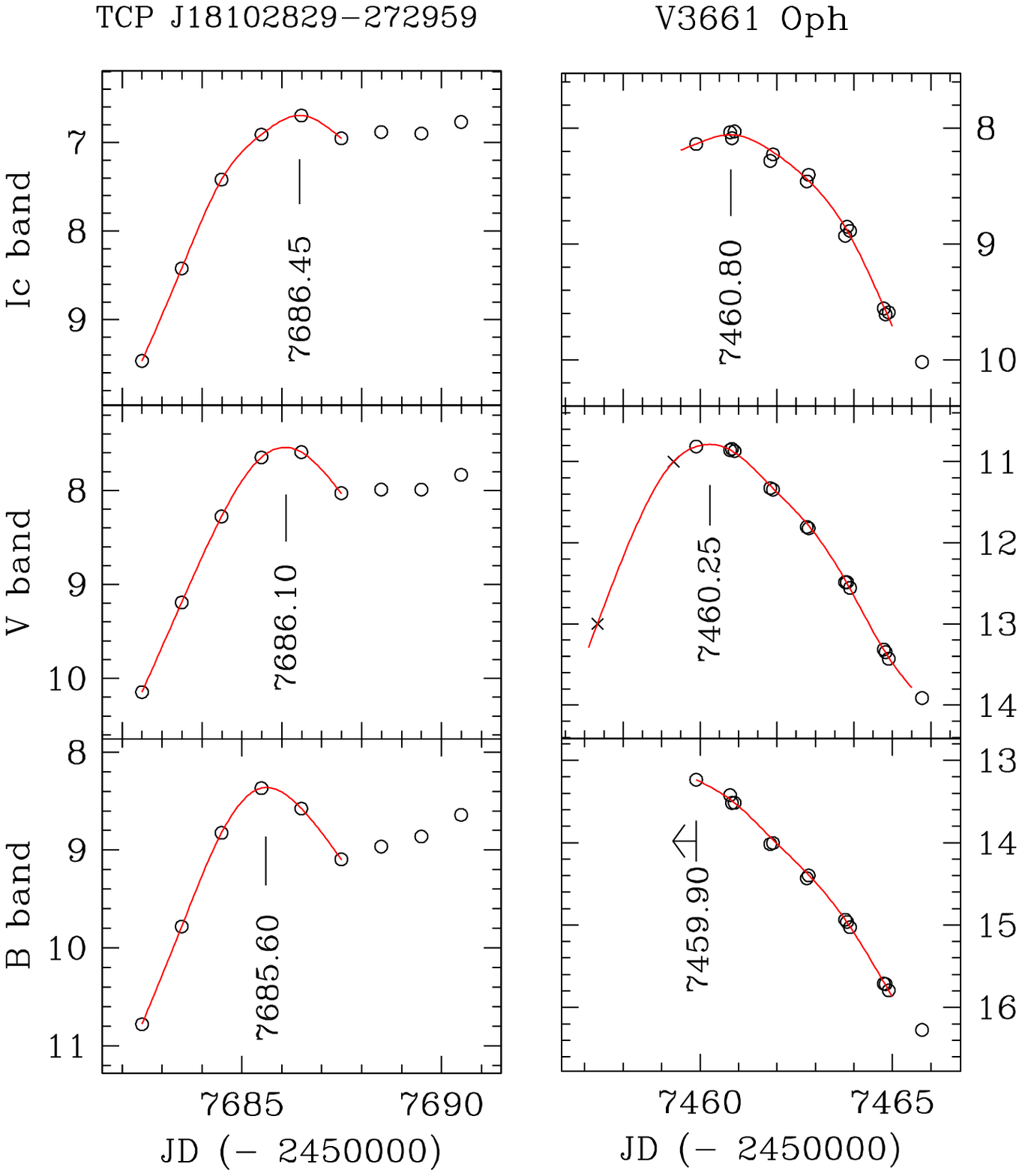}
    \caption{Wavelength dependence of the time of maximum brightness
    in novae TCP J18102829-2729590 and V3661 Oph.}
    \end{figure}

The flux density emitted by the ejecta of the nova expanding as an 
homogeneous, ionized plasma is:
\begin{equation}
f_{\nu} = B_{\nu} (d/D)^2 (1-e^{-\tau_{\nu}})
\end{equation}
where $B_{\nu}$ is the Planck function, $d$ is the linear dimension of the
ejecta that scales as $\sim$$v_{\rm ej}$$(t-t_\circ)$, $D$ is the distance
to the nova and $\tau_{\nu}$ is the free-free optical depth from
bremsstrahlung of electrons.  Following Altenhoff et al.  (1960) and Mezger
\& Henderson (1967), $\tau_{\nu}$ goes as
\begin{equation}
\tau_{\nu} \approx 0.08235 T_{e}^{-1.35} \nu^{-2.1} \int N_{e}^{2}dl
\end{equation}
where $N_e$ and $T_e$ are the electron density (cm$^{-3}$) and temperature
(K), $\nu$ is the frequency in units of 10$^9$ Hz, and the emission measure
$\int N_{e}^{2}dl$ is in pc cm$^{-6}$. The time delay between maximum
brightness reached in different photometric bands can be then expressed as
\begin{eqnarray}
t_{max}^V - t_{max}^B &=&  0.35\times\Theta {\rm ~~~~~(days)}\\
t_{max}^R - t_{max}^V &=&  0.25\times\Theta  \nonumber\\
t_{max}^I - t_{max}^V &=&  0.70\times\Theta  \nonumber\\
t_{max}^J - t_{max}^V &=&  1.55\times\Theta  \nonumber\\
t_{max}^H - t_{max}^V &=&  2.15\times\Theta  \nonumber\\
t_{max}^K - t_{max}^V &=&  2.90\times\Theta  \nonumber
\end{eqnarray}
with
\begin{equation}
\Theta =  \left( \frac{T_{\rm e}}{10^4 {\rm ~K}} \right)^{-0.27}
          \left( \frac{M_{\rm ej}}{10^{-4} {\rm ~M}_\odot} \right)^{+0.4}
          \left( \frac{v_{\rm ej}}{1000 {\rm ~km/sec}} \right)^{-1} 
\end{equation}
where $M_{\rm ej}$ is the ejected mass and $v_{\rm ej}$ the ejection
velocity.  This time delay is the same reason responsible for the maximum
thermal radio emission to be reached $\sim$years past optical maximum
(Hjellming 1974).  The timings given in Figure~14 correspond to $t_{max}^I -
t_{max}^B$=0.85 and $\geq$0.90 for TCP J18102829-2729590 and V3661 Oph,
respectively.  This is close to what expected from Eq.(4) for typical values
of $T_e$, $M_{\rm ej}$ and $v_{\rm ej}$ adopted in computing $\Theta$.

    \begin{figure*}
    \includegraphics[angle=270,width=178mm]{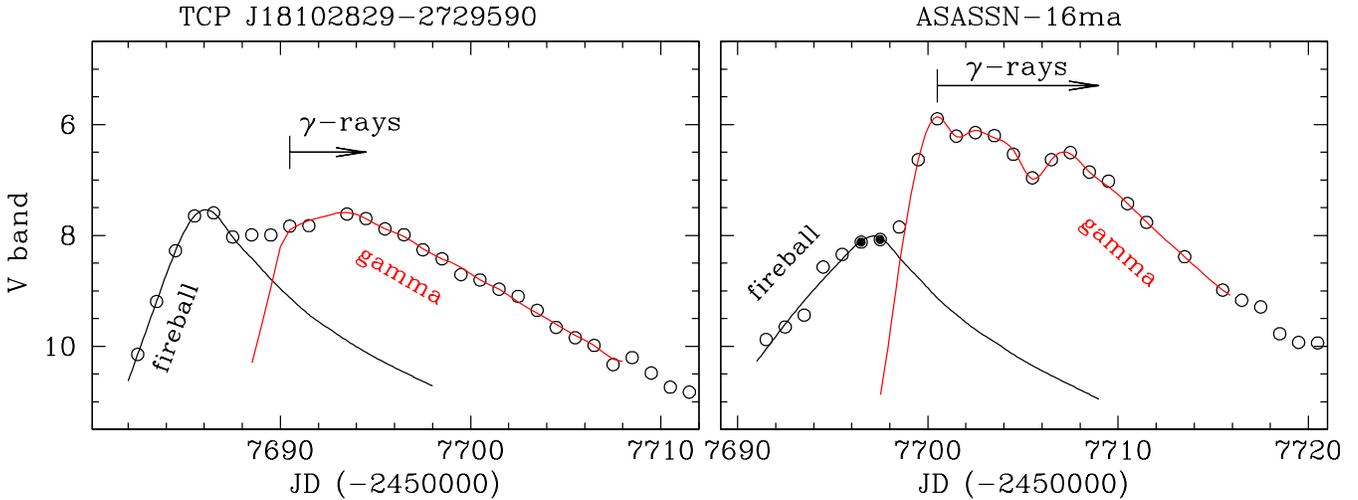}
    \caption{Deconvolution of the lightcurve of the two program Bulge novae
    detected by Fermi-LAT.  The {\em fireball} component is the one associated 
    with free ballistic expansion of ejecta (see sect. 5).  The {\em gamma} 
    component appears and evolves in parallel with the emission detected in
    $\gamma$-rays (see sect. 6). The filled dots are the same as in Figure~12 
    The dip around JD=2457705 in the gamma component for ASASSN-16ma corresponds 
    to a similar dip in the $\gamma$-ray flux recorded by Fermi-LAT (cf. Li et al.  
    2016).}
    \end{figure*}

\section{A second lightcurve component paralleling the emission in
$\gamma$-rays}

The lightcurve of the two $\gamma$-ray program novae, TCP J18102829-2729590
and ASASSN-16ma, is characterized by the distinct presence of two
components, which are highlighted in Figure~15.  The initial or {\em
fireball} component produces a passage through maximum that is dependent on
wavelength as described in the previous section.  The second component
appears at a later time and peaks simultaneously with the detection of the
nova in $\gamma$-rays (for which reason we term it {\em gamma}) and gives
origin to a second maximum which is not wavelength dependent.

The {\em gamma} component of the optical lightcurve behaves synchronously
with the emission observed in $\gamma$-rays.  The preliminary analysis by Li
et al.  (2016) of the daily averaged $\gamma$-ray behavior of ASASSN-16ma,
shows a sudden detection coincident with peak flux on November 8
(JD=2457700.5) followed by a general decline along the following nine days,
with a significant 1-day $\gamma$-ray flux dip observed on November 13
(JD=2457705.5).  The {\em gamma} component of the optical lightcurve in
Figure~15 presents exactly the same behavior: a maximum on November 8 and
a general decline for the following nine days with a 1-day brightness dip
centered on November 13.  Not only the shapes but also the flux ratios
behaved in parallel.  In fact, during the nine days of general decline, the
$\gamma$-ray flux changed by a factor of 3, from 9.7 ($\pm$1.3) to 3.4
($\pm$2.1) $\times$10$^{-7}$ ph cm$^{-2}$ s$^{-1}$ (Li et al.  2016), and
over the same period of time the flux through the $V$-band also declined by
a factor of 3, from $V$=5.89 to $V$=7.02.  Once the daily $\gamma$-ray
behavior of TCP J18102829-2729590 will become available, it will be
interesting to explore if a similar degree of parallelism with the {\em
gamma} component of its optical lightcurve was followed too.

As a further evidence of the link between the $\gamma$-ray emission and the
{\em gamma} component of the lightcurve, it is worth noticing that the
reported mean $\gamma$-ray flux of ASASSN-16ma (Li and Chomiuk 2016) is
2.5$\times$ higher than for TCP J18102829-2729590 (Li et al.  2016).  Well,
the reddening corrected mean flux of the {\em gamma} component of the two
novae in Figure~15 is exactly in the same 2.5$\times$ ratio, or
$<(V)_\circ>$=5.19 and $<(V)_\circ>$=6.15 for ASASSN-16ma and TCP
J18102829-2729590, respectively.

A difference of 2.5$\times$ in the mean $\gamma$-ray flux for the two
program novae, both belonging to the Bulge, seems to disprove the common
belief (eg.  Ackermann et al.  2014), that ($i$) the intrinsic $\gamma$-ray
brightness is similar among normal novae, ($ii$) they can be detected by
Fermi-LAT over only limited distances from the Sun, and therefore ($iii$)
$\gamma$-ray emission is a widespread (if not general) property of novae. 
Judging from ASASSN-16ma and TCP J18102829-2729590, it appears instead that
novae can be firmly detected by Fermi-LAT up to and beyond the Galactic
Bulge, and their intrinsic brightness in $\gamma$-rays can differ greatly. 
Combining this with the relatively low number of normal novae detected to
date by Fermi-LAT (6 novae in total have been detected in $\gamma$-rays, in
contrast to the 69 discovered optically in the same period, cf Morris et al. 
2017), it is tempting to conclude that $\gamma$-ray emission is {\em not} a
wide-spread property for them.

The two-component lightcurve here described for the program $\gamma$-ray
novae brings to mind the two-component ejecta adopted to model the
radio-interferometric observations of some recent novae (Chomiuk et al. 
2014, Weston et al.  2016), a faster polar wind collides with a slower (and
pre-existing ?) equatorial density enhancement.  This scenario applies
however to radio observations extending for months past the initial
eruption.  The second or {\em gamma} component of the optical lightcurve of
program $\gamma$-ray novae develops instead within a few days of the initial
{\em fireball} component, and could therefore trace something different in
the kinematical and geometric arrangment of the ejecta.  We postpone to a
future paper a quantitative modeling of our two-component lightcurve for
$\gamma$-ray novae to include similar data for additional objects and
therefore reinforce the statistics.

\section{Progenitors}

At the position of the four program novae V2949 Oph, V3661 Oph, TCP
J18102829-2729590, and ASASSN-16ma no progenitor is visible in deep OGLE
$I$-band images or DSS plates, which set the minimal outburst amplitudes
listed in Table~3.  For all of them, a progenitor containing a giant or a
sub-giant companion, would have been brighter in $K$ band than the
completeness limit of 2MASS survey in the respective areas, suggesting their
donor star is a dwarf.  For the remaining three program novae (V1534 Sco,
V1535 Sco and MASTER OT J010603.18-744715.8) a progenitor has been proposed
based on positional coincidence with pre-outburst surveys.  We consider in
turn these three novae.

    \begin{figure}
    \includegraphics[width=84mm]{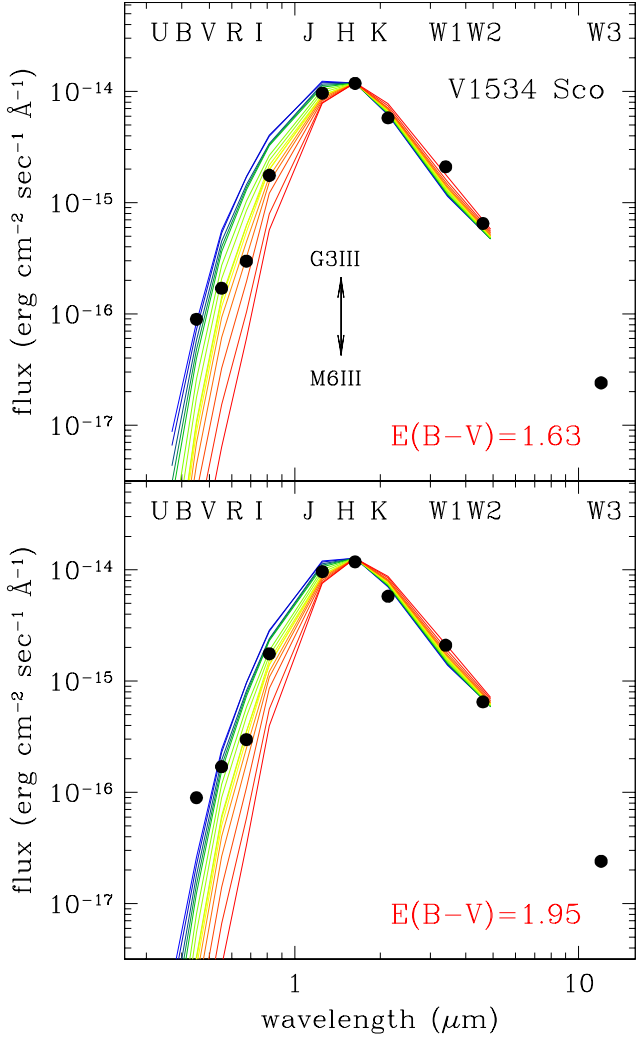}
    \caption{The spectral energy distributions (black dots) of the
     progenitor of nova V1534 Sco.  The families of curves plot the energy
     distribution of giant stars reddened by $E_{B-V}$=1.63 and 1.95,
     corresponding to the extinction from 3D maps of Schlegel, Finkbeiner \&
     Davis (1998) and Schlafly \& Finkbeiner (2011), respectively.  The
     giants go from G3III to M6III, and are scaled to fit the progenitor at
     $H$ band.  Their spectral energy distributions are constructed
     combining data from Koornneef (1983), Bessell (1990), and Fluks et al. 
     (1994).  The best fit ($B$ and $V$ band excluded, see sect 7.1) is for
     an M3III in the upper panel and M1III in the lower one.}
    \end{figure}

    \begin{figure}
    \includegraphics[width=84mm]{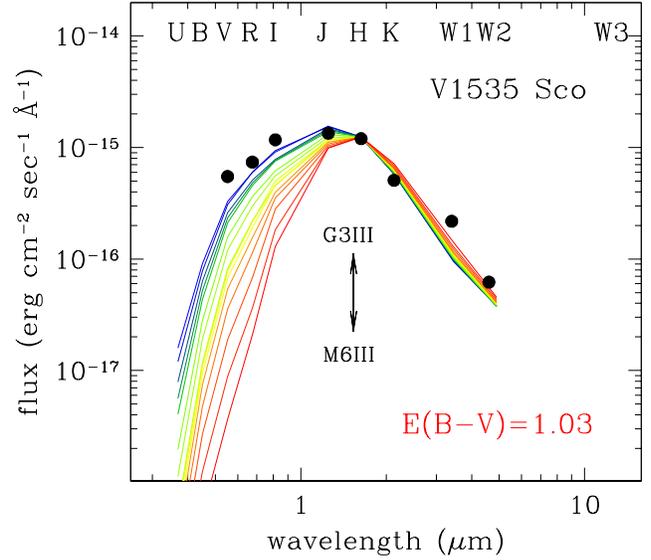}
    \caption{The spectral energy distributions (black dots) of the
     progenitor of nova V1535 Sco. The families of curves
     plot the energy distribution of giant stars reddened by $E_{B-V}$=1.03.
     All details as for Figure~16.}
    \end{figure}

\subsection{V1534 Sco}

Joschi et al.  (2014) proposed 2MASS 17154687-3128303 as the progenitor of
nova V1534 Sco.  At $J$=11.255($\pm$0.042), $H$=10.049($\pm$0.039), and
$K$=9.578($\pm$0.035), it lies at 0.6 arcsec from the position of the nova
reported by SIMBAD.  By fitting with a black-body only its 2MASS and WISE
infrared energy distribution, Joschi et al.  (2014) classified the star 
as an M5III giant, reddened by $E_{B-V}$=0.9.

In Figure~16 we present the observed spectral energy distribution (SED) of
the progenitor of V1534 Sco.  To the 2MASS $J$$H$$K$ and WISE
$W_1$$W_2$$W_3$ infrared data considered by Joschi et al.  (2014), we add
$I$=14.22 mag from DENIS and $R$=17.20 mag from SuperCOSMOS catalogues.  We
have not been able to find quiescence $B$ and $V$ data.  As noted above in
sect.  4.2.1, at latest stages the lightcurve of V1534 Sco became
completely flat, with asymptotic values $B$$\sim$19.6, $V$$\sim$18.3, and
$I$$\sim$14.3.  The latter is practically identical to the pre-outburst
DENIS $I$=14.22 mag value, suggesting that these asymptotic values could be
viable proxies for the brightness in quiescence.  We therefore added the
asymptotic $B$ and $V$ values to the SED of Figure~16.  There we over-plot
to the nova the SEDs of G3III-M6III giants, compiling their optical/IR
intrinsic colors from Koornneef (1983), Bessell (1990), and Fluks et al. 
(1994).  The SEDs of giants are reddened according to the total extinction
along the line of sight to V1534 Sco (cf.  Table~3) as derived from 3D maps
of SFD98 and SF11.  We have already seen in sect.  4.2.1 how these values
for the extinction lead to a correct distance to the nova.  They have been
transformed into the corresponding $E_{B-V}$ and $A_\lambda$ following the
relations calibrated by Fiorucci \& Munari (2003) for M-type giants.

The best fit to $RIJHKW_1W_2W_3$ data in Figure~16 is obtained with an M3III
for $E_{B-V}$=1.63 and an M1III for $E_{B-V}$=1.95.  Overall, the fit with
the M3III is somewhat better.  This is minimally dependent on $KW_1W_2W_3$
bands, while $RIJ$ are far more relevant.  The fit with the M3III provides a
distance of 8.2 kpc, while that with the M1III drops down to 5.0. 
Considering the partnership of the nova with the Bulge, we conclude that the
progenitor of Nova V1534 Sco is well represented by an M3III cool giant
reddened by $E_{B-V}$=1.63.

The $B$ and $V$ points lie above both fit attempts in Figure~16.  There are
at least three suitable explanations for this: (1) the asymptotic $B$ and
$V$ values are still influenced by emission from the nova ejecta, (2) the
severe crowding which fooled the derivation in sect.  4.2 of $E_{B-V}$ from
nova photometry is affecting the $B$ and $V$ brightness of the progenitor
too, and (3) ionization of the cool giant wind by the WD produces extra-flux
at $B$ and $V$ wavelengths.  It is in fact well known how the UBV colors of
symbiotic stars are much bluer than those of the M giants they harbour (cf. 
the UBVRI-JHKL photometric surveys of known symbiotic stars by Munari et al. 
1992 and Henden \& Munari 2008), because of the contribution at shorter
wavelengths by the emission from circumstellar ionized gas.

\subsection{V1535 Sco}

Srivastava et al. (2015) proposed 2MASS 17032617-3504178 ($J$=13.40,
$H$=12.53, and $K$=12.22) as the progenitor.  This star is positionally
concident to better than 0.1 arcsec with the nova, with a Gaia Data Release
1 source of $G$=14.392 mag, and a DENIS counterpart with $I$=15.24 mag.

In Figure~17 we plot the observed SED for the progenitor of V1535 Sco,
combining 2MASS $J$$H$$K$ and WISE $W1$$W2$ infrared data, to which we have
added $I$=15.24 mag from DENIS, $V$=17.05 mag from the YB6 Catalog (USNO,
unpublished; accessed via Vizier at CDS) and $R$=16.33 from SuperCOSMOS.  We
have considered the fit with the same family of energy distributions of
G3III-M6III giants already used in Figure~16 for V1534 Sco, this time
reddened by the same $E_{B-V}$=1.03 derived and discussed above for the
nova.  The fit is clearly unsatisfactory at optical wavelengths, implying
quite blue intrinsic colors for the progenitor.  At the distance given in
Table~3 for the nova, the absolute magnitude of the progenitor would be
M(K)=$-$2.9, which is that expected for a K3-4III giant.  Such a
classification was one of the alternatives (the other being an M4-5III)
considered by Srivastava et al.  (2015).

Although rare, the symbiotic stars with K giants account for $\sim$10\% of
the total in the catalog by Belczy{\'n}ski et al.  (2000). The optical
colors of some of them (cf Munari et al.  1992, Henden \& Munari 2008)
are strongly affected by the blue emission of the K giant wind ionized by
the radiation from the WD companion, and this could easily be a viable
interpretation for the progenitor of V1535 Sco.

\subsection{MASTER OT J010603.18-744715.8}

Mroz et al.  (2016) reported the progenitor was clearly visible in OGLE-IV
survey images at equatorial coordinates R.A.=01:06:03.27, Decl. 
=$-$74:47:15.8 (J2000.0), $I$=20.84 mean magnitude and $V$$-$$I$=+0.16
color.  They add that it showed semi-regular variability on a timescale of
20-30 days.  The blue color reflects into the non-detections by 2MASS and
WISE infrared surveys.

At a distance of 1.03 arcsec from the OGLE position there is a GALEX source
of magnitudes FUV=20.529($\pm$0.331) and NUV=20.573($\pm$0.205), the second
closest GALEX source being 30 arcsec away.  The astrometric proximity and
compatible magnitudes and colors, suggest that the OGLE and GALEX sources
are the same star, of blue colors consistent with those of a disk-dominated
source.

Adopting the $E_{B-V}$=0.08 reddening and 61 kpc distance to SMC listed by
Mateo (1998), the absolute magnitude of the progenitor is $M(V)$=$+$1.8,
which suggests a sub-giant as the donor star.  A giant of the T~CrB type
would shine at $M_V$$\sim$$-$0.5 (Sowell et al.  2007), while the mean
magnitude for novae with dwarf companions is $M_V$$\sim$4.5 (Warner 1995).
The presence of a sub-giant is consistent with the non-detection of the
progenitor during the 2MASS survey.

\section{Acknowledgements} We would like to thank S. Dallaporta, F. 
Castellani, G.  Alsini and R.  Belligoli for some check observations 
carried out on the program targets.


\begin{thebibliography}{}
\bibitem[\protect\citeauthoryear{Ackermann et al.}{2014}]{2014Sci...345..554A} Ackermann M., et al., 2014, Sci, 345, 554 
\bibitem[\protect\citeauthoryear{Altenhoff et al.}{1960}]{Altenhoff} Altenhoff W., Mezger P.~G., Strassl H., Wendker H., Westerhout G., 1960, Veroff Sternwarte Bonn 59, 48
\bibitem[\protect\citeauthoryear{Ayani}{2015}]{2015IAUC.9279....3A} Ayani K., 2015, IAUC, 9279, 3 
\bibitem[\protect\citeauthoryear{Ayani \& Maeno}{2014}]{2014CBET.3841....1A} Ayani, K., Maeno, S., 2014, CBET, 3841, 1
\bibitem[\protect\citeauthoryear{banerjee}{2012}]{b73} Banerjee D.P.K., Ashok, N.M., 2012, BASI, 40, 243
\bibitem[\protect\citeauthoryear{Bessell}{1990}]{1990PASP..102.1181B} Bessell M.~S., 1990, PASP, 102, 1181 
\bibitem[\protect\citeauthoryear{Belczy{\'n}ski et al.}{2000}]{2000A&AS..146..407B} Belczy{\'n}ski K., Miko{\l}ajewska J., Munari U., Ivison R.~J., Friedjung M., 2000, A\&AS, 146, 407 
\bibitem[\protect\citeauthoryear{Bode \& Evans}{2012}]{2012clno.book.....B} Bode M.~F., Evans A., 2012, eds., Classical Novae, Cambridge University Press
\bibitem[\protect\citeauthoryear{Buscombe \& de Vaucouleurs}{1955}]{1955Obs....75..170B} Buscombe W., de Vaucouleurs G., 1955, Obs, 75, 170 
\bibitem[\protect\citeauthoryear{Campbell et al.}{2015}]{2015ATel.8155....1C} Campbell H., et al., 2015, ATel, 8155,  
\bibitem[\protect\citeauthoryear{Capaccioli et al.}{1989}]{1989AJ.....97.1622C} Capaccioli M., della Valle M., Rosino L., D'Onofrio M., 1989, AJ, 97, 1622 
\bibitem[\protect\citeauthoryear{Chomiuk et al.}{2014}]{2014Natur.514..339C} Chomiuk L., et al., 2014, Nature, 514, 339 
\bibitem[\protect\citeauthoryear{Chomiuk et al.}{2016}]{2016ATel.8841....1C} Chomiuk L., Strader J., Stanek K.~Z., Kochanek C.~S., Holoien T.~W.-S., Shappee B.~J., Prieto J.~L., Dong S., 2016, ATel, 8841,  
\bibitem[\protect\citeauthoryear{Cousins}{1980}]{1980SAAOC...1..166C} Cousins A.~W.~J., 1980, SAAOC, 1, 166 
\bibitem[\protect\citeauthoryear{Downes \& Duerbeck}{2000}]{2000AJ....120.2007D} Downes R.~A., Duerbeck H.~W., 2000, AJ, 120, 2007 
\bibitem[\protect\citeauthoryear{Duerbeck}{2008}]{Duerbeck} Duerbeck H.~W., 2008, in Classical Novae, M.-F. Bode and Evans eds., Cambridge University Press, pag. 1
\bibitem[\protect\citeauthoryear{Fiorucci \& Munari}{2003}]{2003A&A...401..781F} Fiorucci M., Munari U., 2003, A\&A, 401, 781 
\bibitem[\protect\citeauthoryear{Fitzpatrick}{1999}]{1999PASP..111...63F} Fitzpatrick E.~L., 1999, PASP, 111, 63 
\bibitem[\protect\citeauthoryear{Fluks et al.}{1994}]{1994A&AS..105..311F} Fluks M.~A., Plez B., The P.~S., de Winter D., Westerlund B.~E., Steenman H.~C., 1994, A\&AS, 105, 311
\bibitem[\protect\citeauthoryear{Frank et al.}{2016}]{2016ATel.8817....1F} Frank S., Wagner R.~M., Starrfield S., Woodward C.~E., Neric M., 2016, ATel, 8817,  
\bibitem[\protect\citeauthoryear{Graham}{1982}]{1982PASP...94..244G} Graham J.~A., 1982, PASP, 94, 244 
\bibitem[\protect\citeauthoryear{Henden et al.}{2012}]{2012JAVSO..40..430H} Henden A.~A., Levine S.~E., Terrell D., Smith T.~C., Welch D., 2012, JAVSO, 40, 430 
\bibitem[\protect\citeauthoryear{Henden \& Munari}{2008}]{2008BaltA..17..293H} Henden A., Munari U., 2008, BaltA, 17, 293 
\bibitem[\protect\citeauthoryear{Henden \& Munari}{2014}]{2014CoSka..43..174M} Henden A., Munari U., 2014, in Observing Techniques, Instrumentation and Science for Meter-Class Telescopes, T. Pribulla ed., CoSka, 43, 518 
\bibitem[\protect\citeauthoryear{Hjellming}{1974}]{1974gegr.book..159H} Hjellming R.~M., 1974, in Galactic and Extra-Galactic Radio Astronomy, G.L. Verschuur and K.I. Kellermann eds., Springer-Verlag New York, pag. 159
\bibitem[\protect\citeauthoryear{Jablonski \& Oliveira}{2016}]{2016ATel.9684....1J} Jablonski F., Oliveira A., 2016, ATel, 9684,  
\bibitem[\protect\citeauthoryear{Jelinek et al.}{2014}]{2014ATel.6025....1J} Jelinek M., Cunniffe R., Castro-Tirado A.~J., Rabaza O., Hudec R., 2014, ATel, 6025,  
\bibitem[\protect\citeauthoryear{Johnson \& Morgan}{1953}]{1953ApJ...117..313J} Johnson H.~L., Morgan W.~W., 1953, ApJ, 117, 313
\bibitem[\protect\citeauthoryear{Joshi et al.}{2014}]{2014ATel.6032....1J} Joshi V., Banerjee D.~P.~K., Venkataraman V., Ashok N.~M., 2014, ATel, 6032,  
\bibitem[\protect\citeauthoryear{Joshi et al.}{2015}]{2015MNRAS.452.3696J} Joshi V., Banerjee D.~P.~K., Ashok N.~M., Venkataraman V., Walter F.~M., 2015, MNRAS, 452, 3696 
\bibitem[\protect\citeauthoryear{Koornneef}{1983}]{1983A&A...128...84K} Koornneef J., 1983, A\&A, 128, 84 
\bibitem[\protect\citeauthoryear{Krautter}{2008}]{2008ASPC..401..139K} Krautter J., 2008, ASPC, 401, 139 
\bibitem[\protect\citeauthoryear{Kuin et al.}{2016}]{2016ATel.9635....1K} Kuin N.~P.~M., Page K.~L., Williams S.~C., Darnley M.~J., Shore S.~N., Walter F.~M., 2016, ATel, 9635,  
\bibitem[\protect\citeauthoryear{Kuulkers et al.}{2014}]{2014ATel.6015....1K} Kuulkers E., Page K.~L., Saxton R.~D., Ness J.-U., Kuin N.~P., Osborne J.~P., 2014, ATel, 6015,  
\bibitem[\protect\citeauthoryear{Landolt}{1992}]{1992AJ....104..340L} Landolt A.~U., 1992, AJ, 104, 340 
\bibitem[\protect\citeauthoryear{Landolt}{2009}]{2009AJ....137.4186L} Landolt A.~U., 2009, AJ, 137, 4186 
\bibitem[\protect\citeauthoryear{Li \& Chomiuk}{2016}]{2016ATel.9699....1L} Li K.-L., Chomiuk L., 2016, ATel, 9699,  
\bibitem[\protect\citeauthoryear{Li, Chomiuk, \& Strader}{2016}]{2016ATel.9736....1L} Li K.-L., Chomiuk L., Strader J., 2016, ATel, 9736,  
\bibitem[\protect\citeauthoryear{Li et al.}{2016}]{2016ATel.9771....1L} Li K.-L., Chomiuk L., Strader J., Cheung C.~C., Jean P., Shore S.~N., Fermi Large Area Telescope Collaboration, 2016, ATel, 9771,  
\bibitem[\protect\citeauthoryear{Linford et al.}{2015}]{2015ATel.7194....1L} Linford J., et al., 2015, ATel, 7194,  
\bibitem[\protect\citeauthoryear{Lipunov et al.}{2016}]{2016ATel.9631....1L} Lipunov V., et al., 2016, ATel, 9631,  
\bibitem[\protect\citeauthoryear{Littlefield \& Garnavich}{2015}]{2015ATel.8156....1L} Littlefield C., Garnavich P., 2015, ATel, 8156,  
\bibitem[\protect\citeauthoryear{Luckas}{2016}]{2016ATel.9678....1L} Luckas P., 2016, ATel, 9678,  
\bibitem[\protect\citeauthoryear{Lukas}{2016}]{2016ATel.9658....1L} Lukas P., 2016, ATel, 9658,  
\bibitem[\protect\citeauthoryear{McLaughlin}{1960}]{1960stat.conf..585M} McLaughlin D.~B., 1960, in Stellar Atmospheres. J. L. Greenstein ed., University of Chicago Press, 585 
\bibitem[\protect\citeauthoryear{Mateo}{1998}]{1998ARA&A..36..435M} Mateo M.~L., 1998, ARA\&A, 36, 435
\bibitem[\protect\citeauthoryear{Mezger \& Henderson}{1967}]{1967ApJ...147..471M} Mezger P.~G., Henderson A.~P., 1967, ApJ, 147, 471
\bibitem[\protect\citeauthoryear{Morris et al.}{2017}]{2017MNRAS.465.1218M} Morris P.~J., Cotter G., Brown A.~M., Chadwick P.~M., 2017, MNRAS, 465, 1218
\bibitem[\protect\citeauthoryear{Mroz \& Udalski}{2016a}]{2016ATel.8811....1M} Mroz P., Udalski A., 2016a, ATel, 8811,  
\bibitem[\protect\citeauthoryear{Mroz \& Udalski}{2016b}]{2016ATel.9622....1M} Mroz P., Udalski A., 2016b, ATel, 9622,  
\bibitem[\protect\citeauthoryear{Mroz et al.}{2016}]{2016ATel.9683....1M} Mroz P., Udalski A., Pietrukowicz P., 2016, ATel, 9683,  
\bibitem[\protect\citeauthoryear{Munari et al.}{1992}]{1992A&AS...93..383M} Munari U., Yudin B.~F., Taranova O.~G., Massone G., Marang F., Roberts G., Winkler H., Whitelock P.~A., 1992, A\&AS, 93, 383 
\bibitem[\protect\citeauthoryear{Munari et al.}{2011}]{2011MNRAS.410L..52M} Munari U., et al., 2011, MNRAS, 410, L52 
\bibitem[\protect\citeauthoryear{Munari et al.}{2012}]{2012BaltA..21...13M} Munari U., et al., 2012, BaltA, 21, 13 
\bibitem[\protect\citeauthoryear{Munari \& Moretti}{2012}]{2012BaltA..21...22M} Munari U., Moretti S., 2012, BaltA, 21, 22 
\bibitem[\protect\citeauthoryear{Munari et al.}{2013a}]{2013MNRAS.435..771M} Munari U., Dallaporta S., Castellani F., Valisa P., Frigo A., Chomiuk L., Ribeiro V.~A.~R.~M., 2013a, MNRAS, 435, 771 
\bibitem[\protect\citeauthoryear{Munari}{2014}]{2014ASPC..490..183M} Munari U., 2014, in Stella Novae: Past and Future Decades, P. A. Woudt and V. A. R. M. Ribeiro eds., ASP Conf. Ser. 490, 183
\bibitem[\protect\citeauthoryear{Munari et al.}{2014a}]{2014JAD....20....4M} Munari U., Henden A., Frigo A., Dallaporta S., 2014a, JAD, 20,  
\bibitem[\protect\citeauthoryear{Munari et al.}{2014b}]{2014AJ....148...81M} Munari U., et al., 2014b, AJ, 148, 81 
\bibitem[\protect\citeauthoryear{Munari et al.}{2015}]{2015NewA...40...28M} Munari U., Maitan A., Moretti S., Tomaselli S., 2015, NewA, 40, 28
\bibitem[\protect\citeauthoryear{Munari et al.}{2016a}]{2016IBVS.6162....1M} Munari U., Walter F.~M., Hambsch F.-J., Frigo A., 2016a, IBVS, 6162, 1 
\bibitem[\protect\citeauthoryear{Munari et al.}{2016b}]{2016IAUC.9280....2M} Munari U., Sollecchia U., Hambsch F.-J., Frigo A., 2016b, IAUC, 9280,  
\bibitem[\protect\citeauthoryear{Muneer \& Anupama}{2016}]{2016ATel.8853....1M} Muneer S., Anupama G.~C., 2016, ATel, 8853,  
\bibitem[\protect\citeauthoryear{Muneer, Anupama, \& Raveendran}{2015}]{2015ATel.7161....1M} Muneer S., Anupama G.~C., Raveendran A.~V., 2015, ATel, 7161,  
\bibitem[\protect\citeauthoryear{Nelson et al.}{2012}]{2012ATel.4590....1N} Nelson T., Mukai K., Sokoloski J., Chomiuk L., Rupen M., Mioduszewski A., Page K., Osborne J., 2012, ATel, 4590,  
\bibitem[\protect\citeauthoryear{Nelson et al.}{2015}]{2015ATel.7085....1N} Nelson T., et al., 2015, ATel, 7085,  
\bibitem[\protect\citeauthoryear{Ness}{2012}]{2012BASI...40..353N} Ness J.~U., 2012, BASI, 40, 353 
\bibitem[\protect\citeauthoryear{Orio et al.}{2016a}]{2016ATel.9810....1O} Orio M., Behar E., Rauch T., Zemk P., 2016a, ATel, 9810,
\bibitem[\protect\citeauthoryear{Orio et al.}{2016b}]{2016ATel.9970....1O} Orio M., Rauch T., Zemko P., Behar E., 2016b, ATel, 9970,
\bibitem[\protect\citeauthoryear{Osborne et al.}{2010}]{2010ATel.2442....1O} Osborne J.~P., et al., 2010, ATel, 2442
\bibitem[\protect\citeauthoryear{Page et al.}{2013}]{2013ATel.4845....1P} Page K.~L., et al., 2013, ATel, 4845,  
\bibitem[\protect\citeauthoryear{Page et al.}{2016}]{2016ATel.9733....1P} Page K., Osborne J., Kuin P., Shore S., Williams S., Darnley M.~J., 2016, ATel, 9733,  
\bibitem[\protect\citeauthoryear{Page, Osborne, \& Kuulkers}{2014}]{2014ATel.6035....1P} Page K.~L., Osborne J.~P., Kuulkers E., 2014, ATel, 6035,  
\bibitem[\protect\citeauthoryear{Rudy, Crawford, \& Russell}{2016}]{2016ATel.9849....1R} Rudy R.~J., Crawford K.~B., Russell R.~W., 2016, ATel, 9849, 
\bibitem[\protect\citeauthoryear{Saito et al.}{2016}]{2016ATel.9680....1S} Saito R.~K., Minniti D., Catelan M., Angeloni R., 2016, ATel, 9680,  
\bibitem[\protect\citeauthoryear{Schlafly \& Finkbeiner}{2011}]{2011ApJ...737..103S} Schlafly E.~F., Finkbeiner D.~P., 2011, ApJ, 737, 103 (SF11)
\bibitem[\protect\citeauthoryear{Schlegel, Finkbeiner, \& Davis}{1998}]{1998ApJ...500..525S} Schlegel D.~J., Finkbeiner D.~P., Davis M., 1998, ApJ, 500, 525 (SFD98)
\bibitem[\protect\citeauthoryear{Schwarz et al.}{2011}]{2011ApJS..197...31S} Schwarz G.~J., et al., 2011, ApJS, 197, 31 
\bibitem[\protect\citeauthoryear{Seitter}{1990}]{1990LNP...369...79S} Seitter W.~C., 1990, in Physics of Classical Novae, A. Cassatella and R. Viotti eds., Springer-Verlag, Berlin, pag. 79
\bibitem[\protect\citeauthoryear{Shumkov et al.}{2016}]{2016ATel.9621....1S} Shumkov V., et al., 2016, ATel, 9621,  
\bibitem[\protect\citeauthoryear{Smith et al.}{2002}]{2002AJ....123.2121S} Smith J.~A., et al., 2002, AJ, 123, 2121
\bibitem[\protect\citeauthoryear{Soker \& Livio}{1989}]{1989ApJ...339..268S} Soker N., Livio M., 1989, ApJ, 339, 268 
\bibitem[\protect\citeauthoryear{Sowell et al.}{2007}]{2007AJ....134.1089S} Sowell J.~R., Trippe M., Caballero-Nieves S.~M., Houk N., 2007, AJ, 134, 1089 
\bibitem[\protect\citeauthoryear{Srivastava et al.}{2015a}]{2015ATel.7236....1S} Srivastava M., Ashok N.~M., Banerjee D.~P.~K., Venkataraman V., 2015a, ATel, 7236,  
\bibitem[\protect\citeauthoryear{Srivastava et al.}{2015b}]{2015MNRAS.454.1297S} Srivastava M.~K., Ashok N.~M., Banerjee D.~P.~K., Sand D., 2015b, MNRAS, 454, 1297 
\bibitem[\protect\citeauthoryear{Srivastava et al.}{2016}]{2016ATel.8809....1S} Srivastava M., Joshi V., Banerjee D.~P.~K., Ashok N.~M., 2016, ATel, 8809,  
\bibitem[\protect\citeauthoryear{Stanek et al.}{2016}]{2016ATel.9669....1S} Stanek K.~Z., et al., 2016, ATel, 9669,  
\bibitem[\protect\citeauthoryear{Strai{\v z}ys}{1992}]{1992msp..book.....S} Strai{\v z}ys V., 1992, Multicolor Stellar Photometry, Pachart Publishing House (Tucson)
\bibitem[\protect\citeauthoryear{van den Bergh \& Younger}{1987}]{1987A&AS...70..125V} van den Bergh S., Younger P.~F., 1987, A\&AS, 70, 125 
\bibitem[\protect\citeauthoryear{Walter et al.}{2012}]{2012PASP..124.1057W} Walter F.~M., Battisti A., Towers S.~E., Bond H.~E., Stringfellow G.~S., 2012, PASP, 124, 1057 
\bibitem[\protect\citeauthoryear{Walter}{2015}]{2015ATel.7060....1W} Walter F., 2015, ATel, 7060,  
\bibitem[\protect\citeauthoryear{Warner}{1995}]{1995CAS....28.....W} Warner B., 1995, Cataclysmic Variable Stars, Cambridge Univ. Press
\bibitem[\protect\citeauthoryear{Weston et al.}{2016}]{2016MNRAS.457..887W} Weston J.~H.~S., et al., 2016, MNRAS, 457, 887
\bibitem[\protect\citeauthoryear{Williams}{1992}]{b131} Williams R.E., 1992, AJ, 104, 725
\bibitem[\protect\citeauthoryear{Williams \& Darnley}{2016a}]{2016ATel.9628....1W} Williams S.~C., Darnley M.~J., 2016a, ATel, 9628,  
\bibitem[\protect\citeauthoryear{Williams \& Darnley}{2016b}]{2016ATel.9688....1W} Williams S.~C., Darnley M.~J., 2016b, ATel, 9688,  
\bibitem[\protect\citeauthoryear{Woudt \& Ribeiro}{2014}]{2014ASPC..490.....W} Woudt P.~A., Ribeiro V.~A.~R.~M., 2014, eds., Stella Novae: Past and Future Decades, ASP Conf. Ser. 490  
\end{thebibliography}
\end{document}